\documentclass[
reprint,
showkeys,
longbibliography,
nofootinbib,
superscriptaddress,
aps,
notitlepage
]{revtex4-1}
\usepackage[utf8]{inputenc}
\usepackage{mathtools}
\usepackage{amsmath}
\usepackage{amssymb}
\usepackage[colorlinks,unicode]{hyperref}
\usepackage[capitalise]{cleveref}
\usepackage[dvipsnames]{xcolor}
\usepackage{graphicx}
\usepackage{multirow}
\usepackage{enumitem}
\usepackage{bm}
\usepackage{physics}
\usepackage{siunitx}
\makeatletter
\renewcommand\onecolumngrid{%
\do@columngrid{one}{\@ne}%
\def\set@footnotewidth{\onecolumngrid}%
\def\footnoterule{\kern-6pt\hrule width 1.5in\kern6pt}%
}
\makeatother

\newcommand\myshade{80}
\colorlet{mylinkcolor}{ForestGreen}
\colorlet{mycitecolor}{Red}
\colorlet{myurlcolor}{violet}
\usepackage[dvipsnames]{xcolor}

\hypersetup{
  linkcolor  = mylinkcolor!\myshade!black,
  citecolor  = mycitecolor!\myshade!black,
  urlcolor   = myurlcolor!\myshade!black,
  colorlinks = true
}

\newcommand{\revision}[1]{{#1}}              

\DeclareSIUnit\solarmass{\ensuremath{\mathrm{M}_\odot}}
\DeclareSIUnit\parsec{pc}
\DeclareSIUnit\year{yr}

\newcommand{\GRAPPA}{Gravitation Astroparticle Physics Amsterdam (GRAPPA),\\ Institute for Theoretical Physics Amsterdam and Delta Institute for Theoretical Physics,\\ University of Amsterdam, Science Park 904, 1098 XH Amsterdam, The Netherlands} 

\newcommand{\UdeM}{Département de Physique, Université de Montréal, 1375 Avenue Thérèse-Lavoie-Roux, Montréal, QC H2V 0B3, Canada}

\newcommand{\Mila}{Mila -- Quebec AI Institute, 6666 St-Urbain, \#200, Montreal, QC, H2S 3H1}

\newcommand{\IFCA}{Instituto de F\'isica de Cantabria (IFCA, UC-CSIC), Av.~de Los Castros s/n, 39005 Santander, Spain}

\newcommand{\UVA}{Department of Physics, University of Virginia, P.O.~Box 400714, Charlottesville, Virginia 22904-4714, USA}

\newcommand{\IFT}{Instituto de F\'isica Te\'orica UAM-CSIC,  Campus de Cantoblanco, E-28049 Madrid, Spain}

\newcommand{\IFIC}{Instituto de F\'isica Corpuscular, Universidad de Valencia and CSIC, Edificio Institutos de Investigac\'ion, Calle Catedr\'atico Jos\'e Beltr\'an 2, 46980 Paterna, Spain}

\newcommand{\Torino}{Dipartimento di Fisica, Universit\`a di Torino, via P. Giuria 1, I–10125 Torino, Italy}

\newcommand{\rmsp}{\mathrm{sp}} 
\newcommand{\rmDM}{\mathrm{DM}} 
\newcommand{\rmIn}{\mathrm{in}} 
 
\newcommand{\rmsh}{\mathrm{sh}}

\newcommand{\calE}{\mathcal{E}}
\newcommand{\hyp}{\operatorname{{}_2F_1}}

\begin{document}

\preprint{}

\title{Measuring the dark matter environments of black hole binaries with gravitational waves}

\author{Adam Coogan}
\email{adam.coogan@umontreal.ca}
\affiliation{\GRAPPA}
\affiliation{\UdeM}
\affiliation{\Mila}

\author{Gianfranco Bertone}
\email{g.bertone@uva.nl}
\affiliation{\GRAPPA}

\author{Daniele Gaggero}
\email{daniele.gaggero@uam.es}
\affiliation{\IFT}
\affiliation{\Torino}
\affiliation{\IFIC}

\author{\\Bradley J. Kavanagh}
\email{kavanagh@ifca.unican.es}
\affiliation{\IFCA}

\author{David A. Nichols}%
\email{david.nichols@virginia.edu}
\affiliation{\UVA}

\begin{abstract}
    Large dark matter overdensities can form around black holes of astrophysical and primordial origin as they form and grow. This ``dark dress'' inevitably affects the dynamical evolution of binary systems, and induces a dephasing in the gravitational waveform that can be probed with future interferometers. In this paper, we introduce a new analytical model to rapidly compute gravitational waveforms in presence of an evolving dark matter distribution. We then present a Bayesian analysis determining when dressed black hole binaries can be distinguished from GR-in-vacuum ones and how well their parameters can be measured, along with how close they must be to be detectable by the planned Laser Interferometer Space Antenna (LISA). We show that LISA can definitively distinguish dark dresses from standard binaries and characterize the dark matter environments around astrophysical and primordial black holes for a wide range of model parameters. Our approach can be generalized to assess the prospects for detecting, classifying, and characterizing other environmental effects in gravitational wave physics. 
\end{abstract}


\maketitle

\section{Introduction}

The nature of the elusive dark matter (DM) that appears to permeate the Universe remains unknown~\cite{Bertone:2004pz}, despite an intense and diverse research program that includes direct detection experiments~\cite{Graham:2015ouw,Schumann:2019eaa}, indirect searches based on astronomical and cosmological data~\cite{Gaskins:2016cha}, and searches at colliders~\cite{Boveia:2018yeb}. The recent direct detection of gravitational waves (GWs)  \cite{Abbott:2016blz,TheLIGOScientific:2017qsa,GBM:2017lvd,LIGOScientific:2018mvr,LIGOScientific:2020ibl} has ushered in a new era for fundamental physics. Present and future experiments such as LIGO/Virgo/KAGRA \cite{LIGOScientific:2019vkc,TheVirgo:2014hva,Akutsu:2018axf}, LISA \cite{AmaroSeoane:2012km,2017arXiv170200786A}, Einstein Telescope \cite{Sathyaprakash:2012jk}, Cosmic Explorer~\cite{Reitze:2019iox}, Pulsar Timing Arrays \cite{2010CQGra..27h4013H,2013CQGra..30v4009K,Hobbs:2013aka,2009arXiv0909.1058J}, and others will soon shed new light on some of the most fundamental questions in particle physics and cosmology~\cite{Barack:2018yly}, and may in particular elucidate the particle nature of dark matter \cite{Bertone:2018xtm,Bertone:2019irm}.

In this paper, we focus on the prospects for detecting and characterizing overdensities of dark matter around compact object binary systems by investigating the associated gravitational radiation. Under the general hypothesis that dark matter is cold and consists of collisionless particles, significant ``spikes" are expected to form around massive compact objects such as supermassive \cite{Gondolo:1999ef,Ullio:2001fb}, intermediate-mass \cite{Bertone:2005hw} and stellar-mass astrophysical black holes (BHs), as well as around hypothetical black holes of primordial origin \cite{Kohri:2014lza,Eroshenko:2016yve,Boucenna:2017ghj}. These DM structures would lead to a significant increase in the annihilation rate of self-annihilating dark matter candidates, and are interesting targets for indirect dark matter searches \cite{Gondolo:1999ef,Gondolo:2000pn,Ullio:2001fb,Bertone:2001jv,Merritt:2002vj,Bertone:2005hw,Merritt:2006mt,Bertone:2005xz,Zhao:2005zr,Bringmann:2009ip,Lacki:2010zf,Eroshenko:2016yve,Boucenna:2017ghj,Adamek:2019gns,Bertone_2019,Carr_2021}.

DM overdensities around black holes could also be detected and studied by measuring how dynamical friction~\cite{Chandrasekhar1943a,Chandrasekhar1943b,Chandrasekhar1943c} affects the orbits of compact object binaries and the emitted GWs. This would manifest as a ``dephasing'' of the GWs emitted by the binary: a gradual change in phase when compared to an equivalent system without DM. In this context, the most promising systems are intermediate mass-ratio inspirals (IMRIs), consisting of a stellar-mass compact object orbiting around a larger intermediate-mass black hole (IMBH)~\cite{Eda:2013gg,Eda:2014kra,Macedo:2013qea,Barausse:2014tra,Barausse:2014pra,Yue:2017iwc,Yue:2019ndw,Hannuksela:2019vip,Edwards:2019tzf,Kavanagh:2020cfn}. 

\begin{figure}[bth!]
    \begin{center}
    \includegraphics[width=0.45\textwidth]{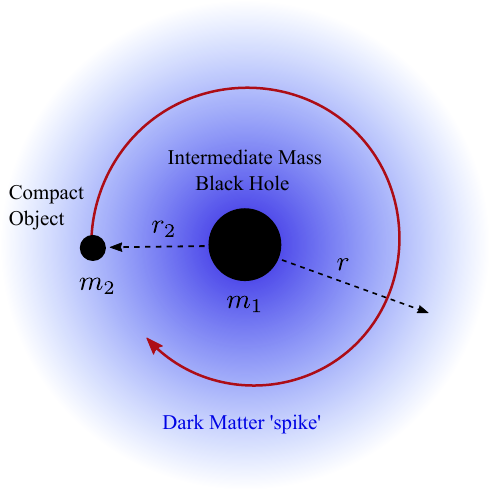}
    \end{center}
    \caption{An intermediate mass black hole (IMBH) of mass $m_1$, surrounded by a ``spike'' of dark matter, is orbited by a lighter compact object $m_2 \ll m_1$ at an orbital radius $r_2$.}
    \label{fig:IMRI}
\end{figure}

We have shown in a previous paper (Ref.~\cite{Kavanagh:2020cfn}, henceforth, Paper I) that in such systems the work done by dynamical friction is typically comparable to (and in some cases much larger than) the total binding energy available in the DM spike. This implies that previous calculations of the DM-induced dephasing, which assumed a non-evolving DM density profile, do not conserve energy and can substantially overestimate the size of the effect. It further highlights the importance of jointly evolving the distribution of DM and the orbital parameters of the binary.

Here, we assess the prospects for detecting and characterizing dark-matter overdensities around black holes with the planned Laser Interferometer Space Antenna (LISA).
The \revision{speed of the} accurate numerical modeling approach described in Paper I is not suitable for a systematic exploration of the model parameter space \revision{requiring tens of thousands of waveform evaluations}. We thus introduce here a new analytical model \revision{that runs $\order{10^5}$ times faster} to approximate gravitational waveforms in the presence of an evolving dark matter distribution.

We then study the detectability of dark dress binaries and present a Bayesian analysis of their discoverability and measurability, defined as follows:
\begin{itemize}
    \item {\bf Detectability.} We call a dark dress \emph{detectable} if it can be detected with LISA. We require in particular that the signal-to-noise ratio with LISA is larger than 15;
    \item {\bf Discoverability.} We call a dark dress \emph{discoverable} if it can be distinguished from a GR-in-vacuum system. To quantify this, we calculate the Bayes factor between the dark dress and vacuum models for the dress's signal;
    \item {\bf Measurability.} We derive full posterior distributions for the dark dress model parameters and demonstrate the feasibility of \emph{measuring} them with LISA in case of detection. 
\end{itemize}

The paper is organised as follows: in \cref{sec:profiles} we describe the properties of dark matter overdensities around intermediate-mass black holes of both astrophysical and primordial origin. In \cref{sec:feedback}, we provide an overview of our approach to the numerical modeling of the dark dress waveform. In \cref{sec:ana_model}, we present an analytical model to approximate gravitational waveforms in the presence of an evolving dark matter distribution. In \cref{sec:framework}, we describe the calculation of the signal-to-noise ratio, and introduce the Bayesian framework to assess discoverability and measurability. In \cref{sec:Results}, we present our results, and in \cref{sec:conclusions}, we discuss them and present our conclusions.

The code used in this work is available at \url{https://github.com/adam-coogan/pydd}.\footnote{
    This excludes the large data files we used to calibrate our approximate phase parametrization (\cref{sec:parametrization}).
}
 
\section{Initial dark matter density profiles}
\label{sec:profiles}

Intermediate-mass black holes (IMBHs) are black holes with mass in the range $10^2 - 10^5\, M_{\odot}$ and can form via a variety of mechanisms, either directly or through accretion and merger of smaller compact objects~\cite{2020ARA&A..58..257G}. Depending on the formation channel, an IMBH can develop a significant dark matter overdensity (``dark dress'') through different physical processes. We review here the main features of such dresses within an astrophysical and a primordial formation scenario, which we use as initial conditions for evolving the system with a binary companion. \revision{We indicate density profiles used as initial data as $\rho_\rmDM(r, t=0)$.}

\subsection{Astrophysical Black Holes}

In the astrophysical scenario, the IMBH progressively grows in a dark matter halo via accretion onto a seed that may originate from a variety of channels, such as the collapse of a stellar-like object formed in a metal-poor environment ({\it ``population III'' star)} \cite{Madau:2001sc}, or {\it direct collapse} of a super-massive star which in turn collapses into a IMBH \cite{Bromm:2002hb}. The DM distribution is altered by the adiabatic growth\footnote{
    Adiabatic growth means that the timescale for the growth of the central BH is much longer than the dynamical timescale for the DM halo.
} of the IMBH and is expected to form a steep \emph{spike} \cite{Gondolo:1999ef,Bertone:2005xz}, with a density profile which is well-described by a power law.

Denoting the IMBH mass by $m_1$, the dark matter distribution for \revision{such a spike}
\begin{equation}
    \rho_\rmDM(r\revision{, t=0}) = \begin{cases}
        \rho_\rmsp \left(\frac{r_\rmsp}{r}\right)^{\gamma_\rmsp} & r_\rmIn \leq r \leq r_\rmsp \\
       0.  & r < r_\rmIn
    \end{cases} \, , \label{eq:rhoDM}
\end{equation}
where $r$ is the distance from the center of the IMBH, $r_\rmsp$ is the size of the spike and $\rho_\rmsp$ is the density at $r_\rmsp$~\cite{Gondolo:1999ef,Bertone:2005hw,Sadeghian:2013laa,Ferrer:2017xwm,PhysRevD.101.024029}. We define the inner radius of the spike as $r_\rmIn = 4 G_N m_1 / c^2$ (twice the Schwarzschild radius), following~\cite{Sadeghian:2013laa}. We will not treat the DM distribution at distances $r > r_\rmsp$, where the length scale $r_\rmsp$ is not a free parameter, but can be expressed as a function of $m_1$, $\rho_\rmsp$ and $\gamma_\rmsp$ as in~\cite{Eda:2014kra} (see also Paper I):
\begin{equation}
\label{eq:r_sp}
    r_\rmsp = \qty[ \frac{(3 - \gamma_\rmsp) 0.2^{3 - \gamma_\rmsp} m_1}{2 \pi \rho_\rmsp} ]^{1/3} \, .
\end{equation}
The spike slope depends on the initial properties of the DM halo where the BH formed. For an initial halo with inner slope $\alpha$ we expect $\gamma_\rmsp = (9 - 2 \alpha) / (4 - \alpha)$~\cite{Gondolo:1999ef}. An initial NFW profile ($\alpha=1$) thus corresponds to $\gamma_\rmsp = 7/ 3 = 2.333\ldots \equiv 2.\overline{3}$. Though we take this model as a benchmark, the formalism we present in this paper can be applied to any DM spike density profile.

A description of the DM spike in terms of $\rho_\rmsp$ and slope $\gamma_\rmsp$ arises naturally in this adiabatic formation model. However, this parametrization is not the most intuitive, because $r_\rmsp$ depends on $\rho_\rmsp$ through \cref{eq:r_sp}. This means that the DM density at a fixed radius scales as $\rho_\rmDM\revision{(r, t=0)} \propto \rho_\rmsp^{(1 -\gamma_\rmsp/3)}$. We therefore introduce the parameter $\rho_6$, such that the DM density in the spike is given by:
\begin{equation}
    \rho_\rmDM(r\revision{, t=0}) = \begin{cases}
        \rho_6 \left(\frac{r_6}{r}\right)^{\gamma_\rmsp} & r_\rmIn \leq r \leq r_\rmsp \\
       0.  & r < r_\rmIn
    \end{cases} \, , \label{eq:rhoDM_6}
\end{equation}
where $r_6 = 10^{-6}\,\mathrm{pc}$ is a fixed reference radius. Comparing \cref{eq:rhoDM} and \cref{eq:rhoDM_6}, we can express $\rho_6$ in terms of $\rho_\rmsp$ as:
\begin{equation}
    \label{eq:rho6-def}
    \rho_6 = \rho_\rmsp^{1- \gamma_\rmsp/3} (k\, m_1)^{\gamma_\rmsp/3} r_6{}^{-\gamma_\rmsp}\,,
\end{equation}
with $k = (3 - \gamma_\rmsp)\,0.2^{3-\gamma_\rmsp}/(2\pi)$. The definition in \cref{eq:rhoDM_6} allows for an intuitive interpretation of $\rho_6$,  with $\rho_\mathrm{DM} \propto \rho_6$ at a fixed radius. We therefore use the parametrization of the spike in terms of ($\rho_6$, $\gamma_\rmsp$) when performing parameter scans.

The density profile in \cref{eq:rhoDM} and~\cref{eq:rhoDM_6} can be generalized by relaxing one or more of the assumptions that went into its derivation:
\begin{itemize}
    \item \emph{Adiabaticity.} The central black hole was assumed to grow adiabatically at the center of the dark matter halo. If the accretion timescale is short compared to the dynamical time of the dark matter halo, or if the black hole is off-center with respect to it, the resulting spike is shallower \cite{Ullio:2001fb}.
    \item \emph{Survival.} The spike is assumed to survive unperturbed after formation. Supermassive black holes (SMBHs) at the centers of galactic halos typically undergo major mergers, which lead to a dramatic suppression of the spike density \cite{Merritt:2002vj}. Furthermore, gravitational interactions with stellar cusps around SMBHs are also expected to deplete spikes \cite{Bertone:2005hw}. IMBHs are more likely to carry unperturbed spikes \cite{Bertone:2005xz} and, as illustrated in Paper I, IMRIs lead to only a minor perturbation of the initial spike after the system has merged.
    \item \emph{Gravity framework.} The density profile was calculated in the framework of Newtonian gravity. A full relativistic treatment leads to a steeper profile, especially in the case of Kerr BHs  \cite{Sadeghian:2013laa,Ferrer_2017}.
    \item \emph{DM interactions.} Dark matter particles are assumed to be cold, collisionless, and non-annihilating. If dark matter is warm, self-interacting, or self-annihilating, the resulting spike is expected to shallower and model-dependent \cite{Gondolo:1999ef,Bertone:2005hw,Shapiro_2016,Hannuksela:2019vip,alvarez2021density}.
\end{itemize}
In our final plots we will indicate the benchmark from Ref.~\cite{Eda:2014kra} of $(m_1, m_2, \rho_\rmsp,\, \gamma_\rmsp) = \qty(\SI{e3}{\solarmass},\, \SI{1.4}{\solarmass}, \SI{226}{\solarmass.\parsec^{-3}},\, 2.\overline{3} )$ (corresponding to $\rho_6 = \SI{5.448e15}{\solarmass / \parsec^3}$), though we emphasize that our treatment can also be applied straightforwardly to the general density profiles described above.

\subsection{Primordial Black Holes}

The intermediate-mass black hole could also be of primordial origin. In this case we also expect the formation of a dark dress. The physical process in this scenario is well-understood: the dress forms due to the nearly-radial infall of cold dark matter in the vicinity of the IMBH, beginning after its formation deep in the radiation era~\cite{Mack:2006gz}.

Analytic calculations based on the theory of secondary infall~\cite{Bertschinger:1985pd}, 1D simulations~\cite{Ricotti:2007jk} and most recently realistic 3D simulations~\cite{Adamek:2019gns} all find that spikes around intermediate-mass PBHs also have power-law density profiles, with $\rho_\rmDM\revision{(r, t=0)} \propto r^{-9/4}$ (see also the recent detailed analysis in Ref.~\cite{Boudaud:2021irr}, which studies spike formation as a function of black hole mass, DM mass and DM kinetic decoupling temperature). Recasting these results into the parametrization \cref{eq:rhoDM}, we find $\rho_s = \SI{1.798e4}{\solarmass.\parsec^{-3}}$, independent of the PBH's mass $m_1$ (corresponding to $\rho_6 = \SI{5.345e15}{\solarmass / \parsec^3}$ for $m_1 = \SI{e3}{\solarmass}$).

We will also highlight this benchmark in our final plots, using the same black hole masses as for the astrophysical one. However, note that since the particles in PBH spikes are moving along extremely elliptical orbits, our modeling assumption that the DM velocity distribution in the spike is isotropic (\cref{sub:dm-spike-evolution}) does not hold, so our results should be interpreted with care.

\section{Numerical dark dress waveform modeling}
\label{sec:feedback}

In this section, we study the evolution of a system consisting of a stellar-mass compact object orbiting around an IMBH surrounded by a dark dress with initial density profile as described in the previous section. The evolution is governed by gravitational wave emission and dynamical friction exerted by the dark dress on the light compact object. We adopt a numerical approach to solve simultaneously for the equation of motion of the binary system and the feedback on the DM spike.

\subsection{Evolution of the IMRI}

As the binary orbits, the orbital energy of the two compact objects is dissipated via gravitational-wave emission and dynamical friction, and the orbital energy evolves as
\begin{equation}
    \label{eq:EnergyBalance}
    \dot E_{\rm orb} = - \dot E_{\rm GW} - \dot E_{\rm DF} \, ,
\end{equation}
where the dot denotes the time derivative $\dv*{t}$. We work in a Newtonian approximation, and we assume the orbit is circular.\footnote{
    See Ref.~\cite{Tang:2020jhx} for a treatment of elliptic orbits, though that analysis is strictly limited to unphysical static DM spikes.
} In this case, for a binary separation $r_2$, the GW dissipation is given by
\begin{equation}
    \label{eq:GWdissipation}
    \dot E_{\rm GW} = \frac{32 G_N^4 M (m_1 m_2)^2}{5 (c r_2)^5} \, , 
\end{equation}
where $m_2$ is the mass of the orbiting compact object and $M = m_1 + m_2$ is the total mass of the binary. Dynamical friction losses are described by~\cite[App.~L]{BinneyAndTremaine}
\begin{equation}
    \label{eq:DFdissipation}
    \dot E_{\rm DF} = 4\pi (G_N m_2)^2 \rho_\rmDM(r_2\revision{, t}) \,\xi\, v^{-1} \log\Lambda \, . 
\end{equation}
The term $\xi$ denotes the fraction of DM particles moving more slowly than the orbital speed (for $\gamma_\rmsp = 7/3$, $m_1=10^3\,M_\odot$, we find $\xi \approx 0.58$, independent of radius). Guided by $N$-body simulations (presented in Paper I), we set the maximum impact parameter for scattered DM particles as the distance where the gravitational force of the orbiting compact object dominates: $b_\mathrm{max} = \sqrt{m_2/m_1}r_2$. This in turn fixes the Coulomb logarithm $\log\Lambda = \log \sqrt{m_1/m_2}$. Combining \crefrange{eq:EnergyBalance}{eq:DFdissipation}, we can determine the evolution of the orbital frequency and phase. The density at the point $r_2$, $\rho_\rmDM(r_2\revision{, t})$, evolves with the binary, using the procedure that we describe next.

\subsection{Evolution of the dark matter density}
\label{sub:dm-spike-evolution}

We have shown in Paper I that the energy dissipated through dynamical friction can be much larger than the binding energy associated with the DM spike. It is therefore necessary to take into account in our modeling the energy transferred into the DM spike.

To this aim, we studied in Paper I the physics of dynamical friction in IMRI systems, and introduced a novel semi-analytic prescription to evolve self-consistently the binary and the dark matter profile, based on the following assumptions:
\begin{enumerate}[label=(\alph*)]
    \item The orbital properties evolve slowly compared to the orbital period. This allows us to consider the rate of energy being injected into the halo as constant over a small number of orbits.
 
    \item The DM halo relaxes to an equilibrium configuration on a short timescale compared to the evolution of the orbital period.
    Therefore, we may update the equilibrium density profile of the DM ``instantaneously'' after energy is injected.

    \item The DM halo is spherically symmetric and isotropic, and remains so throughout the evolution of the system. This allows for a simpler description of the halo, because we only need to model the evolution of the energy of the DM particles and not their angular momentum.
\end{enumerate}

Under these hypotheses, we can describe the DM in the spike with an equilibrium phase space distribution function $f = m_\rmDM \dd[6]{N}/\dd[3]{\mathbf{r}}\,\dd[3]{\mathbf{v}}$. Because of (c), $f$ only depends on the relative energy per unit mass \revision{and point in time}: $f = f(\calE\revision{, t})$, where
\begin{equation}
    \calE(r,v) = \Psi(r) - \frac{1}{2}v^2\,.
\end{equation}
Here, $\Psi(r) = \Phi_0 - \Phi(r)$ is the relative potential, with $\Phi(r)$ the standard gravitational potential and $\Phi_0$ a reference potential. Particles with $\calE > 0$ are considered to be gravitationally bound.  Since the orbital separations we are interested in lie well within the sphere of influence of the central IMBH, we write $\Psi(r) = G_N m_1/r$, neglecting the gravitational potential of  the DM halo. The halo therefore evolves in a fixed gravitational potential, greatly simplifying the calculation. Starting from a given density profile $\rho(r)$, we can construct the distribution function $f(\calE)$ using the Eddington inversion procedure~\cite[p.~290]{BinneyAndTremaine}. 

From assumption (a), we can write the change in the distribution function as $\Delta f \approx T_\mathrm{orb} \, \pdv*{f}{t}$, with $T_\mathrm{orb} = 2\pi\sqrt{(r_2)^3/(G_N M)}$ being the orbital period. Thus, we obtain:   
\begin{equation}
\begin{split}
    \label{eq:dfdt}
    &T_\mathrm{orb} \frac{\partial f(\calE, t)}{\partial t} = - p_\mathcal{E}f(\mathcal{E}, t)\\
    &+\int \left(\frac{\calE}{\calE - \Delta\calE}\right)^{5/2} f(\calE - \Delta\calE, t)  P_{\calE-\Delta\calE}( \Delta\calE) \dd{\Delta\calE} \, ,
\end{split}    
\end{equation}
where $p_\mathcal{E} = \int P_{\mathcal{E}}(\Delta\mathcal{E}) \dd{\Delta\mathcal{E}}$ is the total probability for a particle of energy $\mathcal{E}$ to scatter gravitationally with the compact object during one orbit. This is obtained by integrating the probability $P_\mathcal{E}(\Delta\mathcal{E})$ that a particle with energy $\mathcal{E}$ scatters with the compact object and gains an energy $\Delta\mathcal{E}$. The first term on the right-hand side of \cref{eq:dfdt} corresponds to particles initially at energy $\mathcal{E}$ which scatter off the compact object to a different energy. The second term corresponds to particles scattering into the energy $\mathcal{E}$ from energies $\mathcal{E} - \Delta\mathcal{E}$ (weighted by a phase space factor $\propto \mathcal{E}^{5/2}$).

The change in energy $\Delta\mathcal{E}$ can be straightforwardly related to the impact parameter $b$ of DM particles passing close to the compact object~\cite[App.~L]{BinneyAndTremaine}. The per-orbit scattering probability $P_\mathcal{E}(\Delta\mathcal{E})$ is then evaluated as the fraction of particles with energy $\mathcal{E}$ located at a perpendicular distance $b$ from the compact object orbit (see Paper I for an analytical expression and further details).

\Cref{eq:dfdt} describes the time evolution of the DM distribution function. The (time-dependent) DM density can then be recovered as:
\begin{align}
    \rho_\rmDM(r\revision{, t}) = 4 \pi \int_0^{v_\mathrm{max}(r)} v^2 f\left( \Psi(r) - \frac{1}{2}v^2\revision{, t} \right) \dd{v} \, ,
\end{align}
where $v_\mathrm{max}(r) = \sqrt{2\Psi(r)}$ is the escape speed at radius $r$. The \texttt{HaloFeedback} code implements the prescription described above. It was developed alongside Paper I and allows us to compute the properties and evolution of the DM overdensity. It is publicly available online at \href{https://github.com/bradkav/HaloFeedback}{https://github.com/bradkav/HaloFeedback}~\cite{HaloFeedback}.

\subsection{Evolution of the binary with halo feedback}
\label{sec:binary_with_feedback}

To solve the full system, we must jointly evolve the DM distribution through \cref{eq:dfdt} with the dissipative dynamics of the binary. For the binary's dynamics, it is convenient to write the energy balance condition, \crefrange{eq:EnergyBalance}{eq:DFdissipation} as a function of $r_2$, by recalling that for circular orbits the orbital energy and velocity are $E_{\rm orb} = - G_N m_1 m_2/(2r_2)$ and $v=\sqrt{G_N M/r_2}$ respectively. We will also replace the static $\rho(r)$ and $\xi$ with functions of time. The result is
\begin{equation}
\begin{split}
    &\dot{r}_2 = - \frac{64\, G_N^3\, M \, m_1\, m_2}{5\, c^5\, (r_2)^3} \\ &- \frac{8 \pi\, G_N^{1/2}\, m_2 \, \log\Lambda  r_2^{5/2} \, \rho_\rmDM(r_2,t) \,\xi(r_2, t)}{\sqrt{M} m_1 }  \, .
    \label{eq:r_eom}
\end{split}    
\end{equation}
We start with the static DM spike and the binary at a separation $r_2$ three times larger than the desired $r_2$, so as to determine reasonable initial conditions for the system (as described in Paper I).

Because the evolution of $\rho_\rmDM(r,t)$ depends on $r_2$, we simultaneously evolve \cref{eq:dfdt} and~\cref{eq:r_eom} as a coupled system of partial and ordinary differential equations. Our algorithm to solve the system is the following: first, we evaluate the integrals using Simpson's rule, and then we adopt the method of lines (i.e.\ solving the differential equation on a discretized grid of $\mathcal E$ values). We use a second-order-accurate Runge-Kutta method for numerical integration.

The injection of energy by the inspiraling compact object tends to deplete the DM density at the orbital radius. There is therefore competition between the rate of this depletion (which will reduce the dynamical friction effect) and the inspiraling of the binary towards smaller radii (where the density profile is not yet affected). We note, however, that as DM particles are redistributed from smaller to larger radii during the inspiral, the depletion of the spike is largely transient (see Paper I). A set of animations showing examples of the time evolution of the binary and the profile of the DM spike are available online at \url{https://doi.org/10.6084/m9.figshare.11663676}~\cite{Animations}. 

\section{Analytic dark dress waveform modeling}
\label{sec:ana_model}

The goal of this work is to estimate the detectability of dark dresses and how precisely their parameters can be measured throughout their parameter space. While the numerical modeling approach described up until here is currently the most accurate way to model dark dress waveforms, it is not suitable for this task, requiring $\order{\SI{10}{\hour}}$ \revision{on a single CPU} to track a single system starting at \SI{5}{\year} before coalescence. Here we instead build an analytic approximation of the output of these models. \revision{We begin by reviewing the phase evolution of static dresses. We then explain the qualitative insights from numerical modeling that enable us to construct our approximate model with a similar form. Lastly, we quantitatively connect our insights to the the physics governing dark dress evolution and present our new model's functional form.}

\begin{figure*}[t]
    \centering
    \includegraphics[width=0.36\textwidth]{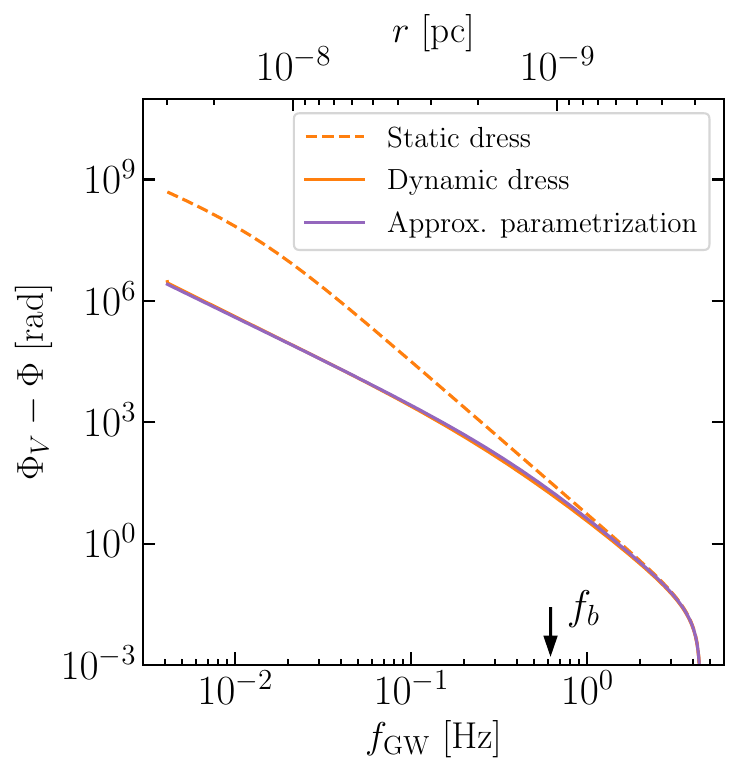}
    \includegraphics[width=0.59\textwidth]{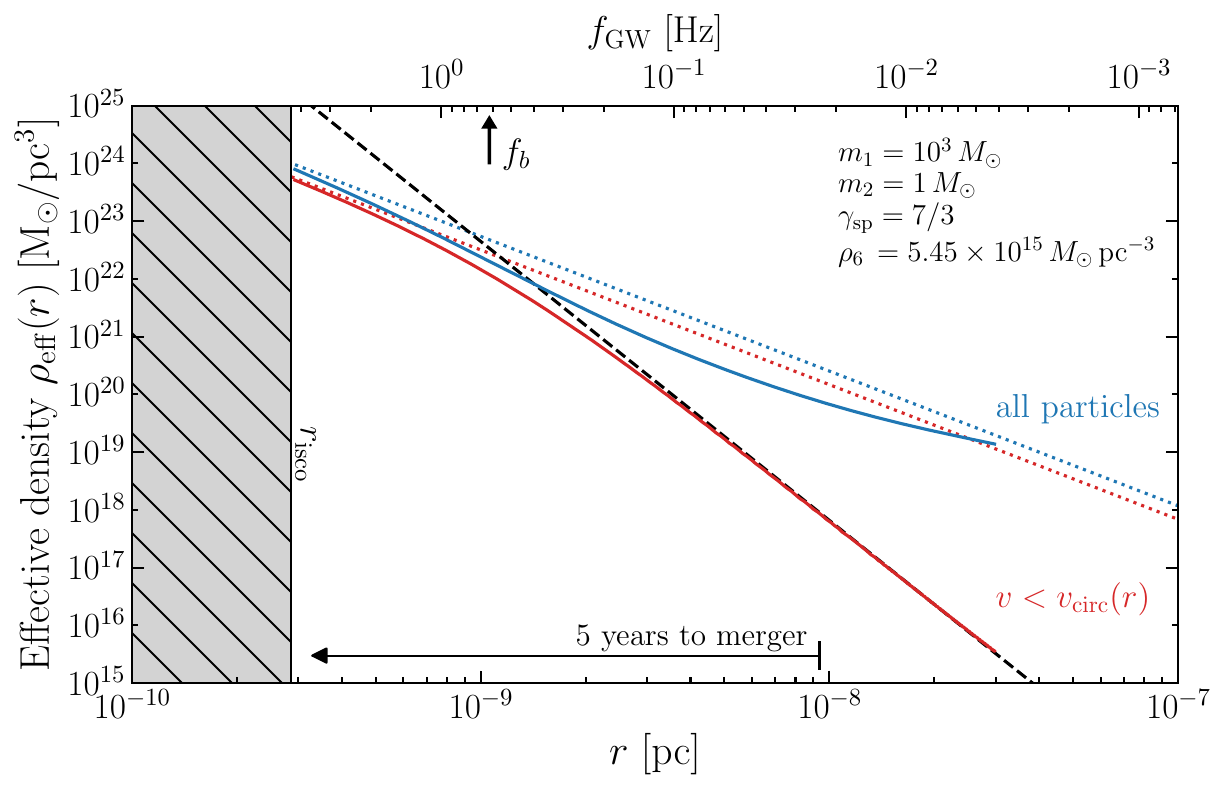}
    \caption{\textbf{Dephasing for a static and dynamic dark dress (left) and the corresponding effective density profile (right).} We assume the benchmark astrophysical system in Table~\ref{tab:benchmarks}. In the left panel, we show the difference in the phase-to-merger between a vacuum inspiral and a system with a static dress (orange dashed) and a dynamic dress (orange solid), using the output from \texttt{HaloFeedback}. For comparison, we also show the dephasing for a dynamic dress using the approximate phase parametrization (purple), described in \cref{sec:parametrization}. In the right panel, we show the DM density at the position of the inspiraling compact object $r = r_2$ in the dynamic case (obtained using \texttt{HaloFeedback}). The blue curve shows the density including all DM particles, while the red curve includes only those particles moving more slowly than the local circular speed. Dotted lines show the unperturbed (static) DM density profile, $\revision{\rho_\rmDM(r, t=0)}$. The dashed black line shows $\rho_\mathrm{eff} \propto \revision{r^{-\gamma_e} \, \rho_\rmDM(r, t=0)}$, with $\gamma_e = 5/2$, as suggested by the shell model at large radii. In both panels, the initial separation of the binary is $r_2 \approx 3 \times 10^{-8}\,\mathrm{pc}$ and an initial period of transient depletion of the DM spike has been removed. The dynamic dress' break frequency $f_b$ is marked by an arrow.}
    \label{fig:dephasing_edp}
\end{figure*}

For a static DM spike (for which $\rho_\rmDM(r, t) = \rho_\rmDM(r\revision{, t=0})$), the phase left until coalescence was derived analytically in Paper I. Assuming circular orbits, the gravitational wave frequency of the quadrupole radiation and the black hole separation are related through $f = \frac{1}{\pi} \sqrt{\frac{G_N M}{r^3}}$. Substituting this into \cref{eq:r_eom}, solving for $f(t)$ and integrating $2\pi$ times the frequency over time from $f$ to the coalescence frequency $f_c$ gives the phase
\begin{equation}
\begin{split}
    &\Phi^\mathrm{S}(f) = \Phi^\mathrm{V}(f) \\
    &\hspace{0.25cm} \times \hyp \qty( 1, \frac{5}{11-2\gamma_\rmsp}, 1 + \frac{5}{11-2\gamma_\rmsp}, - c_f\, f^{-\frac{11 - 2\gamma_\rmsp}{3}} ) \, , \label{eq:sd_phase}
\end{split}    
\end{equation}
defined up to an additive constant $\phi_c$, the ``phase at coalescence''. Here, $\hyp$ is the Gaussian hypergeometric function and
\begin{equation}
    \Phi^\mathrm{V}(f) = \frac{1}{16} \qty(\frac{c^3}{\pi G_N \mathcal{M} f})^{5/3}\,,
\end{equation}
is the phase for a vacuum system with chirp mass $\mathcal{M} = \left(m_{1} m_{2}\right)^{3 / 5}/\left(m_{1}+m_{2}\right)^{1 / 5}$. The DM density profile normalization enters through the parameter $c_f$:\footnote{
    This expression corrects Eq.~(B4) of Paper I, which was missing a factor of 8 in the denominator. This typo was simply one in typesetting the equation; it did not have an impact on any of the results of Paper I.
}
\begin{equation}
    c_f = \frac{5c^5}{8 m_1^2} \pi^{\frac{2(\gamma_\rmsp - 4)}{3}} G_N^{-\frac{2 + \gamma_\rmsp}{3}}  (m_1 + m_2)^{\frac{1 - \gamma_\rmsp}{3}} r_\rmsp^{\gamma_\rmsp} \xi \rho_s \log\Lambda \, . \label{eq:c_f}
\end{equation}
The resulting dephasing $\Delta\Phi^\mathrm{S} \equiv \Phi^\mathrm{V} - \Phi^\mathrm{S}$ is approximately a broken power law. The break occurs at the point where the gravitational wave and dynamical friction energy loss rates are equal ($\sim \SI{0.015}{\hertz}$ for our astrophysical and PBH benchmarks summarized in Table~\ref{tab:benchmarks}):
\begin{equation}
    f_\mathrm{eq} = c_f^{\frac{3}{11 - 2 \gamma_\rmsp}} \, .
\end{equation}
Below and above $f_\mathrm{eq}$ the dephasing can be expanded as
\begin{equation}
    \Delta\Phi^\mathrm{S}(f) = \begin{dcases}
        \Phi^\mathrm{V}(f) \, , & f \ll f_\mathrm{eq}\\
        \frac{5 c_f \, \Phi^\mathrm{V}(f)}{2 (8 - \gamma_\rmsp)} f^{-\frac{11 - 2 \gamma_\rmsp}{3}} \, ,  & f \gg f_\mathrm{eq}
    \end{dcases} \, .
    \label{eq:sd_lims}
\end{equation}
Starting at a large orbital separation, the effects of dynamical friction in a static dress system drastically reduce the number of cycles before coalescence, compared to the vacuum system. This explains why at low frequency the dephasing goes as $\Delta\Phi^\mathrm{S} \equiv \Phi^\mathrm{V} - \Phi^\mathrm{S} \approx \Phi^\mathrm{V}$.

\revision{We similarly} define the phase $\Phi^\mathrm{D}$ for binaries with a dynamic dark dress and a corresponding dephasing $\Delta\Phi^\mathrm{D} \equiv \Phi^\mathrm{V} - \Phi^\mathrm{D}$. While the dephasing $\Delta\Phi^\mathrm{D}$ for dynamic dark dresses is more complicated (and far smaller than for static dark dresses), it can be evaluated using the prescription in \cref{sec:binary_with_feedback}. \revision{As shown in the left panel of \cref{fig:dephasing_edp}, the dynamic dress' dephasing has a similar broken power law form to the dephasing of a static dress, but with a different break frequency and exponents. In the figure and the rest of this work, we denote the dynamic dress dephasing break frequency by $f_b$.}

\revision{The form of the dephasing can be further understood} by studying a quantity we call the \emph{effective density profile} (EDP). From the equations of motion for the dark dress, it is apparent that while the evolution of the whole DM halo is quite complex, only its density at the position of the inspiraling compact object matters for the purposes of computing the evolution of the binary separation. In other words, if we knew \textit{a priori} the density of slow-moving DM particles seen by the inspiraling compact object along its true trajectory 
\begin{equation}
    \rho_\mathrm{eff}(r) \equiv \xi(r_2(t) = r,t) \, \rho_\mathrm{DM}(r_2(t) = r, t)\,,
\end{equation}
we could substitute it for $\xi \, \rho_\rmDM$ in \cref{eq:r_eom} and solve for $r_2(t)$ to derive the true binary separation, treating the EDP as a static halo. While $\rho_\mathrm{eff}(r)$ is of course not known prior to running numerical models, it is useful to study afterwards to build intuition about the dynamics driving the binary separation.

We show an example effective density profile in the right panel of \cref{fig:dephasing_edp}. At small separations (high frequencies) the EDP approaches the initial $\revision{\rho_\rmDM(r, t=0)} \propto r^{-\gamma_\rmsp}$ density profile. For separations larger than the break point $r_b$ (\revision{the separation corresponding to the dephasing break frequency} $f_b$), the EDP falls off according to a steeper power law $r^{-(\gamma_\rmsp + \gamma_e)}$, where $\gamma_e$ is nearly independent of the dark dress's parameters.

Since the EDP is much smaller than the initial dark matter density, the energy loss rate from dynamical friction for a dynamic dress is always much smaller than from gravitational wave emission, as illustrated in \cref{fig:EnergyLoss}. This means the $f \gg f_\mathrm{eq}$ limiting case of \cref{eq:sd_lims} can be used to translate the approximate broken power law EDP into an approximate broken power law dephasing:
\begin{equation}
    \Delta\Phi^\mathrm{D} \propto \begin{dcases}
        \revision{\Phi^\mathrm{V}(f)} \, f^{-\frac{11 - 2 (\gamma_\rmsp + \gamma_e)}{3}} \, , & f \ll f_b\\
        \revision{\Phi^\mathrm{V}(f)} \, f^{-\frac{11 - 2 \gamma_\rmsp}{3}} \, , & f \gg f_b
    \end{dcases} \, .
    \label{eq:dephasing-regimes}
\end{equation}

This qualitative analysis of dynamic dresses raises a physics question and a practical one: where do the dynamic dephasing break frequency $f_b$ and slope $\gamma_e$ of the EDP come from, and what \revision{specific} parametrization should we use for $\Phi^\mathrm{D}$ to carry out our detectability analysis? In the remainder of this section, we derive approximate scaling relations for $f_b$ and $\gamma_e$ that are borne out by the results of numerical modeling, and we construct an analytic model for $\Phi^\mathrm{D}$ that is closely related to $\Phi^\mathrm{S}$.

\subsection{Deriving the effective density profile parameters}

\begin{figure}[t]
    \centering
    \includegraphics[width=0.5\textwidth]{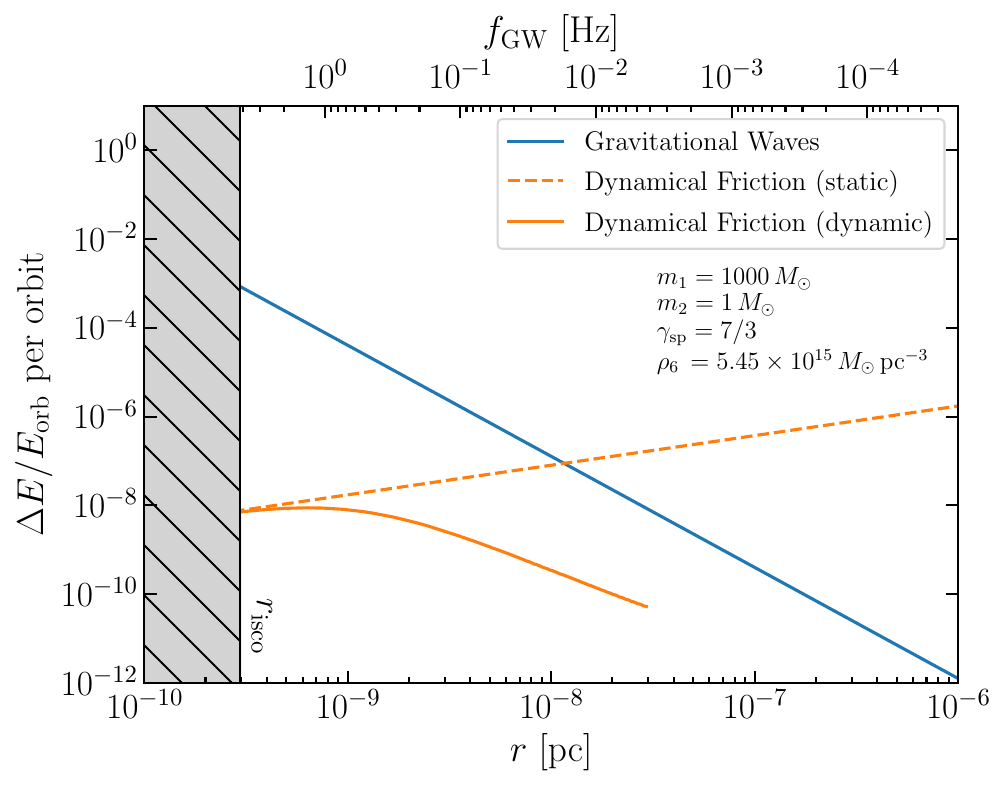}
    \caption{\textbf{Binary energy loss due to gravitational waves and DM dynamical friction.} For the dynamic dress (orange), we use the output from \texttt{HaloFeedback} (starting from $r = 3\times10^{-8}\,\mathrm{pc}$) to determine the effective density profile and calculate the energy loss from \cref{eq:DFdissipation}. A period of transient behavior due to the initial depletion of the DM spike has been removed. For the orbital separations of interest, energy losses due to dynamical friction are always subdominant to gravitational wave emission.}
    \label{fig:EnergyLoss}
\end{figure}

When the DM halo's evolution is neglected, the break $f_\mathrm{eq}$ in the power-law behavior of the dephasing occurs when the timescales for inspiraling due to dynamical friction $t_\mathrm{DF}$ and due to gravitational wave emission $t_\mathrm{GW}$ become equal. In reality, at the separations we consider these timescales never become equal since the DM halo is significantly altered. This is illustrated in \cref{fig:EnergyLoss}, which shows that once the dynamic nature of the halo is taken into account, energy losses due to dynamical friction are always much smaller than those from gravitational wave emission. Instead, the shape of the effective density profile suggests the break in the dynamic dress dephasing $f_b$ occurs when the timescale for \emph{depletion} of the dark matter halo $t_\mathrm{dep}$ at the position of the inspiraling compact object matches the gravitational wave emission timescale. For $f \ll f_b$, we expect $t_\mathrm{dep} \ll t_\mathrm{GW}$ and the halo can be efficiently depleted. For $f \gg f_b$, we expect that the system will inspiral quickly due to GW emission and the behaviour will tend towards that of a static system.

The timescale for GW emission can be estimated straightforwardly from the first term in \cref{eq:r_eom}:
\begin{equation}
\label{eq:t_GW}
    t_\mathrm{GW} \sim \frac{r_2}{\dot{r}_2} \sim \frac{5 c^{5} r_{2}^{4}}{64 G_N^{3} (m_1 + m_2) m_{1} m_{2}} \propto \frac{r_2^4}{m_1^{2} m_2} \, .
\end{equation}
The depletion timescale can be estimated by considering the behaviour of individual DM particles under repeated ``kicks'' from the orbiting compact object. For a particle with energy $\mathcal{E}$, we estimate the depletion timescale as
\begin{equation}
    t_\mathrm{dep}(\mathcal{E}) \sim \, N_\mathrm{req} \, \frac{T_\mathrm{orb}}{p_\mathcal{E}} \, .
\end{equation}
Here, $T_\mathrm{orb}$ is the orbital period of the inspiraling compact object and $p_\mathcal{E}$ is the probability that it scatters with a DM particle of energy $\mathcal{E}$ during a single orbit. The ratio of the two is therefore the typical time between kicks. We also multiply by $N_\mathrm{req}$, the number of kicks required to decrease the particle's energy from $\mathcal{E}$ to $\frac{1}{2}\Psi(r_2) \sim r_2^2 \, v_\mathrm{orb}$. Above this energy, the particle will be moving faster than the compact object $v_\mathrm{orb}(r_2)$ and is therefore considered irrelevant for dynamical friction.\footnote{Recall that the relative specific energy is defined as $\mathcal{E} = \Psi(r) - \frac{1}{2}v^2$.} The required number of kicks is approximately
\begin{equation}
    N_\mathrm{req} \sim \frac{\mathcal{E} - \frac{1}{2}\Psi(r_2)}{\langle \Delta \mathcal{E}\rangle} \, .
\end{equation}
The denominator is the average kick size. The depletion timescale as a function of separation $t_\mathrm{dep}(r)$ is obtained by averaging $t_\mathrm{dep}(\mathcal{E})$ over the phase-space distribution of particles moving more slowly than the orbital speed at $r$. As detailed in \cref{app:f_b}, this yields
\begin{equation}
    t_\mathrm{dep}(r) \sim \frac{m_1^{3/2} r^{3/2}}{m_2^2 \log(1 + m_1 / m_2)} g(\gamma_\rmsp) \, ,
\end{equation}
where
\begin{align}
    g(\gamma_\rmsp) &\equiv \frac{2^{3-\gamma_\rmsp} + \gamma_\rmsp - 4}{(3-\gamma_\rmsp)(2-\gamma_\rmsp) \, h(\gamma_\rmsp)},\\
    h(\gamma_\rmsp) &\equiv \operatorname{B}_{1}\left(\gamma_\rmsp-\frac{1}{2}, \frac{3}{2}\right)-\operatorname{B}_{\frac{1}{2}}\left(\gamma_\rmsp-\frac{1}{2}, \frac{3}{2}\right)\,,
\end{align}
and $\operatorname{B}_x(a,b)$ is the incomplete beta function. Increasing $m_1$ leads to an increase in the depletion timescale, as the DM halo becomes more tightly bound, while increasing $m_2$ shortens the depletion time, as more energy is injected by the orbiting compact object.

Expressing the timescales in terms of the gravitational wave frequency and equating at $f_b$ then gives the scaling relation
\begin{equation}
    f_b \propto \frac{m_2^{3/5}}{m_1^{8/5}} \qty[ \frac{1}{g(\gamma_\rmsp)}\, \log\qty( 1 + \frac{m_1}{m_2} ) ]^{3/5} \, , \label{eq:f_b_scaling_ana}
\end{equation}
independent of the overall density normalization of the DM dress $\rho_\rmsp$. This estimate of the \revision{dynamic dress dephasing} break frequency is relatively simplistic. For example, it does not take into account the fact that as the DM particles gain energy (as $\mathcal{E}$ decreases), their probability of scattering shrinks. In solving the full system, we might therefore expect a slightly different scaling of the break frequency. However, these estimates help illuminate the physics behind the evolution of dynamic DM halos, as well as motivating the parametrization which we present in \cref{sec:parametrization}.

We now turn to estimating $\gamma_e$, the change relative to the initial slope of the effective density profile at large radii, or equivalently at $f \ll f_b$. In this case, the compact object inspirals only very slowly due to the emission of GWs. We will therefore assume that at large radii, there is sufficient time for dynamical friction to act until the DM halo is completely depleted. This means that all of the gravitational binding energy stored in a shell of DM of thickness $\mathrm{d}r$ will be converted into orbital energy as the compact object inspirals from $r +\mathrm{d}r \rightarrow r$. This ``shell model'' for dynamical friction was first presented in Appendix A of Paper I.

Following this shell model, energy balance means that we can write:
\begin{equation}
    \label{eq:EnergyBalance_shellmodel}
    \dot{E}_\mathrm{orb} = -\dot{E}_\mathrm{GW} - \xi \dot{r}_2\frac{\mathrm{d}U_\mathrm{sh}}{\mathrm{d}r_2}\,,
\end{equation}
where the second term on the right gives the rate at which binding energy can be extracted from the DM spike. We include a factor of $\xi$ to account for the fact that only those particles moving more slowly than the local circular speed can be depleted by dynamical friction. For a power-law spike with \revision{an initial} unperturbed density profile $\revision{\rho_\rmDM(r, t=0)}$, the gravitational binding energy of a thin shell of DM at radius $r$ is:
\begin{equation} \label{eq:dUsh_dr}
\begin{split}
    \frac{\mathrm{d} U_{\mathrm{sh}}}{\mathrm{d}r} &\approx -\frac{G_N 
    m_1 m_{\mathrm{DM}}(r)\left(3-\gamma_{\rmsp}\right)}{r^{2}}\\
    &\approx -4 \pi G_N m_1 r \rho_\rmsp\left(\frac{r_\rmsp}{r}\right)^{\gamma_\rmsp}\\
    &= -4 \pi G_N  m_1 r \revision{\rho_\rmDM(r, t=0)}\,,
\end{split}
\end{equation}
where the DM mass enclosed within a radius $r$ is:\footnote{In principle, the spike should be truncated at small radii $r < r_\mathrm{isco}$, but we ignore that small correction here.}
\begin{equation}
    m_{\mathrm{DM}}(r)=\frac{4 \pi \rho_{\rmsp} r_{\rmsp}^{\gamma_{\rmsp}}}{3-\gamma_{\rmsp}} r^{3-\gamma_{\rmsp}}\,.
\end{equation}
Assuming that the inspiral of the compact object is driven predominantly by GW emission (as illustrated in \cref{fig:EnergyLoss}), we have:
\begin{equation} \label{eq:dotr2GW}
    \dot{r}_{2}=-\frac{64 G_N^{3} M m_{1} m_{2}}{5 c^{5}\left(r_{2}\right)^{3}}\,.
\end{equation}
Comparing \cref{eq:EnergyBalance_shellmodel} with the more general energy balance equation, \cref{eq:EnergyBalance}, we can make the identification:
\begin{equation} \label{eq:dotEdf_shell}
\begin{split}
    \dot{E}_\mathrm{DF} &= 4\pi (G_N m_2)^2 \rho_\mathrm{DM}(r_2\revision{, t}) \,\xi\, v^{-1} \log\Lambda\\
    &= \xi \dot{r}_2\frac{\mathrm{d}U_\mathrm{sh}}{\mathrm{d}r_2}\,.
\end{split}
\end{equation}
Here, we will fix the DM density as $\rho_\rmDM(r_2\revision{, t=0}) = \rho_\mathrm{eff}(r_2)$; that is, the dynamical friction energy loss is driven by the (potentially depleted) local DM density as seen by a compact object inspiraling according to \cref{eq:dotr2GW}. We can thus infer the scaling of the effective density profile at large radii:
\begin{equation}
    \rho_\mathrm{eff}(r_2) \propto \revision{\rho_\rmDM(r_2, t=0)} \, r_2^{-5/2} \propto r_2^{-(\gamma_\rmsp + 5/2)}\,.
\end{equation}
From this we can make the identification $\gamma_e = 5/2$. The black dashed line in the right panel of \cref{fig:dephasing_edp} illustrates this expected scaling of the effective density as $r^{-(\gamma_\rmsp + \gamma_e)}$ in the large separation regime. The effective density extracted from running \texttt{HaloFeedback} (solid red line) matches this expected scaling closely for $r \gg r_b$. 

We note that at very large radii, dynamical friction may come to dominate the energy losses of the binary. This occurs when the available binding energy in the DM shell exceeds the energy required to bring the binary to smaller radii: $|\xi \mathrm{d}U_\rmsh/\mathrm{d}r_2| > \mathrm{d}E_\mathrm{orb}/\mathrm{d}r_2$. In this case, the DM is not entirely depleted and the binary may inspiral due to dynamical friction alone, contrary to the assumption in \cref{eq:dotr2GW}. The shell model is therefore no longer valid at large radii, $r_2 > [m_2/(8\pi\xi\rho_\rmsp r_\rmsp^{\gamma_\rmsp})]^{1/(3-\gamma_\rmsp)}$. For the astrophysical benchmark system which we will adopt (see \cref{tab:benchmarks}), the shell model breaks down only for $r_2 \gtrsim 10^{-4}\,\mathrm{pc}$, and we therefore conclude that it should still provide an accurate description over the range of radii we consider.

\subsection{Approximate phase parametrization}
\label{sec:parametrization}

Now that we have motivated that the dark dress dephasing behaves roughly as a broken power law in frequency and have some analytic control over its shape, we must parametrize it with a function. By analogy with static dark dresses, we select the four-parameter family
\begin{equation}
\begin{split}
    &\hat{\Phi}(f) \equiv \Phi^\mathrm{V}(f) \\
    &\hspace{0.25cm} \times \qty{ 1 - \eta \, y^{-\lambda} \qty[ 1 - \hyp\qty( 1,\, \vartheta,\, 1 + \vartheta,\, -y^{-\frac{5}{3 \vartheta}} ) ] } \, ,
    \label{eq:phase_hyp}
\end{split}
\end{equation}
where $y \equiv f / f_t$ is a dimensionless frequency variable. The parameters $f_t$ and $\vartheta$ control the position of the change in power laws and the slope of the power law at high frequencies. For
\begin{equation}
\begin{split}
    \eta &= 1\, ,\\
    \lambda &= 0\, ,\\
    \vartheta &= \frac{5}{11 - 2 \gamma_\rmsp} \, ,\\
    f_t &= f_\mathrm{eq} = c_f^{\frac{3}{11 - 2 \gamma_\rmsp}}\, ,
\end{split}
\end{equation}
this reduces to the static dress's phase, \cref{eq:sd_phase}. A nonzero $\lambda$ value controls the overall power law behavior of the dephasing, and $\eta$ scales the magnitude of the dephasing.

To fit the parameters in $\hat{\Phi}$ in the dynamic case, using \cref{eq:dephasing-regimes} we match the dephasing onto that of a static dress with an effective density profile. Based on our analysis in the previous subsection, we fix $\gamma_e = 5/2$. Taylor-expanding $\hat{\Phi}$ and equating powers of frequency for $f \ll f_b$ fixes $\lambda$. Doing the same in the $f \gg f_b$ limit fixes $\vartheta$ and $\eta$. The result is
\begin{equation}
\begin{split}
    \vartheta &= \frac{5}{2\gamma_e} \, ,\\
    \lambda &= \frac{11 - 2(\gamma_\rmsp+\gamma_e)}{3} \, ,\\
    \eta &= \frac{5 + 2 \gamma_e}{2 (8 - \gamma_\rmsp)} \qty( \frac{f_\mathrm{eq}}{f_b} )^{\frac{11 - 2\gamma_\rmsp}{3}},\\
    f_t &= f_b \, .
\end{split}
\end{equation}

\begin{figure}
    \centering
    \includegraphics[width=0.5\textwidth]{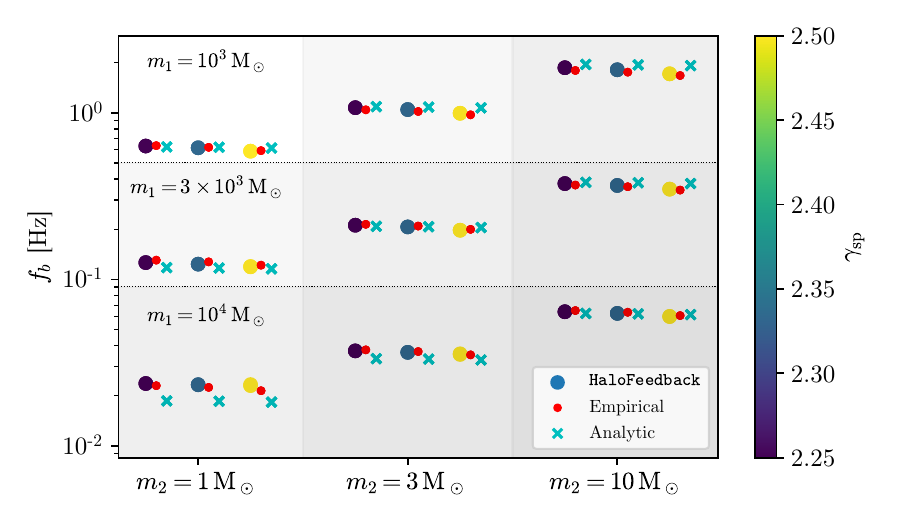}
    \caption{\textbf{Comparison of \revision{dynamic dress dephasing} break frequency values extracted from \texttt{HaloFeedback} waveforms with estimates using our empirical (red) and analytic (cyan) scaling relations.} The systems shown here were used to calibrate the empirical scaling relation (\cref{eq:f_b_scaling_ana}). Each grey box shows three systems with the same values of $(m_1,m_2)$, spaced horizontally to differentiate the distinct values of $\gamma_\rmsp$. The normalization of the analytic scaling relation (\cref{eq:f_b_scaling_ana}) was set to give a good fit to the extracted $f_b$ values.}
    \label{fig:f_b_scaling}
\end{figure}

Instead of using our analytic scaling relation \cref{eq:f_b_scaling_ana} to set $f_b$, we take it as inspiration to fit an empirical relation. For calibrating this empirical relation, we use the last five years before coalescence of the phases of 80 numerical modeling runs generated by the \texttt{HaloFeedback} code.\footnote{
    \revision{Parameter combinations were chosen from $m_1 \in \{ 10^3, \num{3e3}, 10^4 \}\, \si{\solarmass}$, $m_2 \in \{ 1, 3, 10 \}\, \si{\solarmass}$, $\rho_\rmsp \in \{ 20, 200, 2000 \}\, \si{\solarmass.\parsec^3}$ and $\gamma_\rmsp \in \{ 2.25, 2.\bar{3}, 2.5 \}$.}
} We performed nonlinear least-squares fits to find the $f_b$ values that give the best match of our phase approximation to the phase of each run. We find the following function approximates these fit values with an average error of $\sim 2\%$:
\begin{equation}
    f_b = \beta \qty( \frac{m_1}{\SI{1000}{\solarmass}} )^{-\alpha_1} \qty( \frac{m_2}{\si{\solarmass}} )^{\alpha_2} \qty[ 1 + \zeta \log \frac{\gamma_\rmsp}{\gamma_r} ] \, , \label{eq:f_b_scaling}
\end{equation}
where $\alpha_1 = 1.4412$, $\alpha_2 = 0.4511$, $\beta = \SI{0.8163}{\hertz}$, $\zeta = -0.4971$ and $\gamma_r = 1.4396$. \Cref{fig:f_b_scaling} compares this relation and the analytic one with the calibration values of $f_b$. The plot and the numerical values of $\alpha_1$ and $\alpha_2$ show that the analytic scaling relation $f_b \propto m_2^{0.6} / m_1^{1.6}$ overestimates how steeply the break frequency scales with the black hole masses, leading to $\sim 6\%$ error averaged over the calibration systems. To validate this scaling relation, we ran \texttt{HaloFeedback} on 13 additional systems and extracted $f_b$ values as for the calibration systems.\footnote{
    Parameter combinations were randomly sampled from $m_1 \in [10^3, 10^5]\, \si{\solarmass}$, $m_2 \in [1, 100]\, \si{\solarmass}$, $\gamma_\rmsp \in [2.25, 2.5]$ and $\rho_6 \in [\num{e13}, \num{e17}]\, \si{\solarmass/\parsec^3}$, with the mass ratio restricted to $q < 10^{-2.5}$.
} The empirical scaling relation gave an average relative error of $\lesssim 2\%$ for these systems, indicating good performance away from the calibration points.

The hyperparameters $(\alpha_1, \alpha_2, \beta, \rho, \gamma_r)$ depend on how much of the waveform preceding coalescence is used for calibration, which implies that our waveform model should not be extrapolated beyond five years before merger. With this in mind and the additional caveats that it does not exactly capture the shape of the turnover near $f_b$ nor the frequency dependence of the slope at low frequencies, we find our analytic dephasing model provides a good match with the numerically computed dephasing over its range of validity.

\section{Assessing detectability, discoverability and measurability}
\label{sec:framework}

Given the phase as a function of frequency for a binary system it is straightforward to compute the corresponding signal strain, as we review in \cref{app:phase-to-strain}. We use the Newtonian-order strain and average over the polarization, sky position and inclination angles. The angular averages reduce the number of extrinsic parameters, which are not the focus of this study. We assume that IMRIs are distributed uniformly over the sky, and do not expect the localization to depend on the presence of a dark dress. We assume the strain time series measured by a detector $d(t)$ is the sum of the signal $s(t)$ and the detector noise $n(t)$. When the noise $n(t)$ is Gaussian, the likelihood function for this signal given some model waveform $h_{\bm{\theta}}(t)$ with parameters $\bm{\theta}$ is defined (up to a normalizing constant) as
\begin{equation}
\begin{split}
    p\qty( d | h_{\bm{\theta}} ) &\propto \exp\qty[ -\frac{1}{2} \braket{d - h_{\bm{\theta}}} ] \\
    &\propto \exp\qty[ \braket{h_{\bm{\theta}}}{d} - \frac{1}{2} \braket{h_{\bm{\theta}}} ] \, ,
    \label{eq:likelihood}
\end{split}
\end{equation}
where we absorbed a factor independent of $\bm{\theta}$ into the normalizing constant. The noise-weighted inner product is defined using the LISA sensitivity curve $S_n(f)$ (namely, the one-sided power spectral density of the detector noise multiplied by the frequency-dependent response function averaged over sky location and polarization) as
\begin{equation}
    \braket{a}{b} = 4 \Re \int_0^\infty \dd{f} \frac{\tilde{a}(f)^* \, \tilde{b}(f)}{S_n(f)} \, . \label{eq:inner-product}
\end{equation}
The analytic expression for the LISA sensitivity curve that we use is given in Ref.~\cite{Robson:2018ifk}. The model parameters we adopt for the vacuum and dark dress waveforms are $\bm{\theta}_\mathrm{V} = \{ \mathcal{M} \} \cup \bm{\theta}_\mathrm{ext}$ and $\bm{\theta}_\mathrm{D} = \{ \gamma_\rmsp, \rho_6, \mathcal{M}, \log_{10} q \} \cup \bm{\theta}_\mathrm{ext}$ respectively, where $q = m_2/m_1$ is the mass ratio of the binary. The extrinsic parameters $\bm{\theta}_\mathrm{ext}$ are the luminosity distance to the system and the phase and time at coalescence:
\begin{equation}
    \bm{\theta}_\mathrm{ext} \equiv \qty{ D_L, \phi_c, \tilde{t}_c } \, . \label{eq:theta-ext}
\end{equation}

It substantially reduces the computational cost of the analysis to maximize the likelihood with respect to the extrinsic parameters.\footnote{
    A more involved alternative would be to eliminate the extrinsic parameters through marginalization, as explained in App.~C of Ref.~\cite{Thrane_2019}.
} The inner product between the signal and model waveform can be rewritten by making the $\tilde{t}_c$ and $\phi_c$ dependence explicit:
\begin{equation}
    \braket{h_{\bm{\theta}}}{d} = 4 \Re\qty[ e^{i \phi_c} \int_0^\infty \dd{f} \frac{\tilde{h}^*_{\bm{\theta},\phi_c=\tilde{t}_c=0}(f) \, \tilde{d}(f)}{S_n(f)} e^{-2 \pi i f\, \tilde{t}_c} ] \, .
\end{equation}
This shows that the optimization over $\phi_c$ can be performed by replacing the $\Re$ with an absolute value to rotate the integral along the real axis (see, e.g., Sec.~II~B of Ref.~\cite{Owen:1995tm}). Since the integral has the form of a Fourier transform, a single fast Fourier transform gives the value of $\tilde{t}_c$ maximizing the inner product (see, e.g., Sec.~II~A of Ref.~\cite{Owen:1995tm}). Lastly, since $\tilde{d}_L$ enters the likelihood only through the amplitude of $h_{\bm{\theta}}$, it can be maximized over analytically. The resulting maximized likelihood is
\begin{equation}
    p_\mathrm{max}(d | h_{\bm{\theta}}) \equiv \exp\qty[ \frac{\braket{h_{\bm{\theta}}}{d}_\mathrm{max}^2}{2\braket{h_{\bm{\theta}}}} ] \, , \label{eq:pmax}
\end{equation}
where $\braket{\cdot}_\mathrm{max}$ indicates the inner product maximized over $\phi_c$ and $\tilde{t}_c$. We implement the waveform and likelihood calculations using the \texttt{jax}~\cite{jax2018github} Python package.\footnote{
    Since \texttt{jax} does not contain the special function $\hyp$, we interpolate it over a fine grid.
} For computing the match maximized over extrinsic parameters, we use a grid of 100,000 frequencies between the initial frequency and the frequency of the innermost stable circular orbit (ISCO).

To assess the parameters for which dark dresses can be detected, we assume a matched filtering data analysis using a template bank. Such a search requires computing the test statistic $\rho$ between the measured waveform and each template $h_{\bm{\theta}}$ in the bank, defined as
\begin{equation}
    \rho(h_{\bm{\theta}} | d) = \frac{\braket{d}{h_{\bm{\theta}}}}{\sqrt{\braket{h_{\bm{\theta}}}}} \, , \label{eq:matched-filter-ts}
\end{equation}
and finding the template for which it is maximized. For a sufficiently large bank, the expectation value over noise realizations of this quantity approaches the optimal signal-to-noise ratio
\begin{equation}
    \max_{h_{\bm{\theta}}} \rho(h_{\bm{\theta}}|d) \to \mathrm{SNR}(s) = \sqrt{\braket{s}} \, . \label{eq:snr}
\end{equation}
Determining what value of $\rho$ corresponds to a detection requires detailed analysis of the false alarm probability, which depends on factors such as the number of templates in the bank and the observation time. Here we assume systems with an optimal SNR larger than 15 will be detectable at LISA with matched filtering (based on the estimates in, e.g., Ref.~\cite{Moore:2019pke}).

We call a dark dress \emph{discoverable} if it can be distinguished from a GR-in-vacuum system. To quantify this we take a Bayesian approach by computing the Bayes factor for the dark dress and vacuum models for a signal with a dark dress.\footnote{
    An alternative tool for model comparison is to study the ratio of the likelihood maxima for the two models. For nested models, by Wilks' theorem~\cite{Wilks1938}, twice the log of this ratio follows $\chi^2$ distribution. However, Wilks' theorem does not apply since our waveform models are not nested. In particular, when $\rho_6 = 0$, $\gamma_\rmsp$ and $q$ can take on any values without impacting the waveform. While methods exist to determine the correct sampling distribution of the likelihood~\cite{Gross:2010qma,Algeri:2015zpa,2018arXiv180303858A}, they require a substantial number of likelihood evaluations. Another less severe problem in applying Wilks' theorem is that the null hypothesis $\rho_6 = 0$ lies on the boundary of the $\rho_6$ parameter range, which is straightforward to account for~\cite{Chernoff1954,Algeri:2019arh}.
} This is defined as the ratio of the evidences for the signal under each model,
\begin{equation}
    \operatorname{BF}(d) \equiv \frac{p\qty(d | \mathrm{D})}{p\qty(d | \mathrm{V})} \, , \label{eq:bayes-fact}
\end{equation}
where the evidence for a model with parameters $\bm{\theta}$ is
\begin{equation}
    p\qty( d ) = \int \dd{\bm{\theta}} p_\mathrm{max}\qty(d | h_{\bm{\theta}}) \, p(\bm{\theta}) \,, \label{eq:evidence}
\end{equation}
and $p(\bm{\theta})$ is the prior. A signal for which the Bayes factor exceeds 100 can be understood as decisively favoring a dark dress rather than GR-in-vacuum interpretation~\cite{jeffreys1998theory,10.2307/2291091}. We use the nested sampling~\cite{Skilling2004,2019S&C....29..891H} code \texttt{dynesty}~\cite{dynesty} to carry out the evidence calculation. Lastly, since nested sampling also produces posterior probability distributions, we use these to determine how well a dark dress's parameters can be measured.

For simplicity, we ignore the detector noise component of the measured strain, taking $d = s$. This corresponds to replacing the log of the likelihood in \cref{eq:likelihood} with its average over an ensemble of noise realizations. While we expect any given noise realization would lead to a Bayes factor and posteriors biased by a small amount from the $d = s$ case, our approach captures the median behavior of the analysis.

\section{Results}
\label{sec:Results}

\begin{table}
    \centering
    \begin{tabular}{c c c}
        \toprule
        Parameter & Astrophysical & Primordial \\
        \colrule
        $m_1$ [\si{\solarmass}] & \num{e3} & \num{e3}\\
        $m_2$ [\si{\solarmass}] & \num{1.4} & \num{1.4}\\
        $\rho_6$ [\SI{e15}{\solarmass/\parsec^3}] & \num{5.448} & \num{5.345}\\
        $\rho_\rmsp$ [\si{\solarmass/\parsec^3}] & \num{226} & \num{1.798e4}\\
        $\gamma_\rmsp$ & $\num{7/3} = 2.\overline{3}$ & \num{9/4} = \num{2.25}\\
        $D_L$ [\si{\mega\parsec}] & \num{76} & \num{76}\\
        $m_\mathrm{DM}(< \SI{e-6}{\parsec})$ [\si{\solarmass}] & \num{0.102} & \num{0.090}\\
        \botrule
    \end{tabular}
    \caption{\textbf{The dark dress benchmarks whose discoverability and measurability we study.} The rows indicate the black hole masses (defined in the detector frame), initial dark matter halo parameters, luminosity distance and amount of dark matter contained within \SI{e-6}{\parsec}. Assuming a Planck cosmology, the redshift of the systems is \num{0.017}.}
    \label{tab:benchmarks}
\end{table}

Our analysis of the detectability, discoverability and measurability of dark dresses focuses on the astrophysical and primordial black hole benchmarks introduced in \cref{sec:profiles}, whose parameters are given in \cref{tab:benchmarks}. Their masses are defined in the detector frame, and thus related to the source-frame ones through the redshift via $m_\mathrm{det} = m_\mathrm{src} (1 + z)$. We assume LISA measures their signals for five years before the coalescence. This is slightly longer than the nominal mission lifetime of four years but well within the total potential lifetime of ten years~\cite{2017arXiv170200786A}.

The prior impacts the Bayes factor calculation by changing the parameter space volume and affects the posteriors when they impinge on the prior boundary. For the prior on $\gamma_\rmsp$ we use a uniform distribution $\mathcal{U}(2.25, 2.5)$. This is the parameter range expected for an astrophysical dark dress that formed in a DM halo with an initial slope $0 \leq \alpha \leq 2$, roughly the values consistent with simulations (see \cref{sec:profiles}). We also used this range to calibrate our waveform model. We use a uniform prior $\mathcal{U}(0, \SI{2.88e18}{\solarmass/\parsec^3})$ on $\rho_6$, which amply covers the benchmark values and the possibility that the system formed in a substantially denser dark matter environment than expected. The prior on $\log_{10} q$ is set to $\mathcal{U}(-3.5, -2.5)$, corresponding to the range of mass ratios for which we can reliably model the DM halo's evolution and extract the frequency scale $f_b$. Lastly, for the dark dress and vacuum systems we use the same uniform prior on the chirp mass. We take the prior broad enough to encompass the posterior in the $\mathcal{M}$ direction; the precise range does not matter, because it cancels in the dark-dress to GR-in-vacuum evidence ratio in the Bayes factor.\footnote{
    Nested sampling is slow to converge when the prior is much wider than the posterior~\cite{dynesty}. Since we use uniform priors, when necessary we adopt narrow priors to carry out nested sampling that enclose the posterior's support and subsequently rescale the evidence. A rough estimate of the posterior's support was obtained using the Markov Chain Monte Carlo sampler \texttt{emcee}~\cite{2013PASP..125..306F}.
}

\begin{figure}[t]
    \centering
    \includegraphics[width=0.45\textwidth]{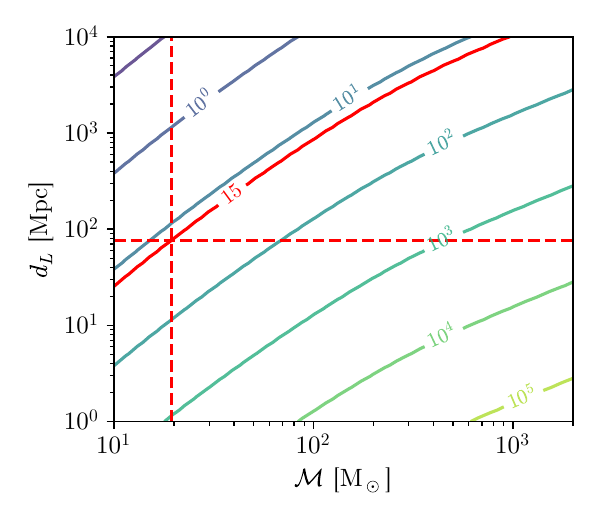}
    \caption{\textbf{Detectability: signal-to-noise ratios for a vacuum binary as a function of chirp mass and luminosity distance.} The solid red contour highlights a reasonable detection threshold for IMRIs~\cite{Moore:2019pke}. The chirp mass and distance of the benchmarks we analyze are indicated by the dashed red lines. As explained in the text, including the effects of the dark dress does not significantly impact this plot.}
    \label{fig:snrs}
\end{figure}

We start by assessing detectability. The signal-to-noise ratio (SNR) of a vacuum binary is plotted in \cref{fig:snrs} as a function of chirp mass and distance. Since the dephasing of a dark dress is quite small relative to its total phase, we found its SNR is very well approximated by the corresponding system without dark matter. In contrast, when the evolution of the dark matter halo is neglected, the SNR falls off steeply for large $\rho_6$ and $\gamma_\rmsp$. This is because the amplitude of the strain scales as $A \propto \ddot{\Phi}^{-1/2} \propto [\dv*{f}{t}]^{-1/2}$ [cf. \cref{eq:amp_plus,eq:amp_times,eq:h0}]. Since the dynamical friction effect is significantly larger for a static dress than a dynamical one, the frequency increases more rapidly with time, leading to a smaller amplitude and consequentially a smaller SNR.

\begin{figure*}[t]
    \centering
    \includegraphics[width=\textwidth]{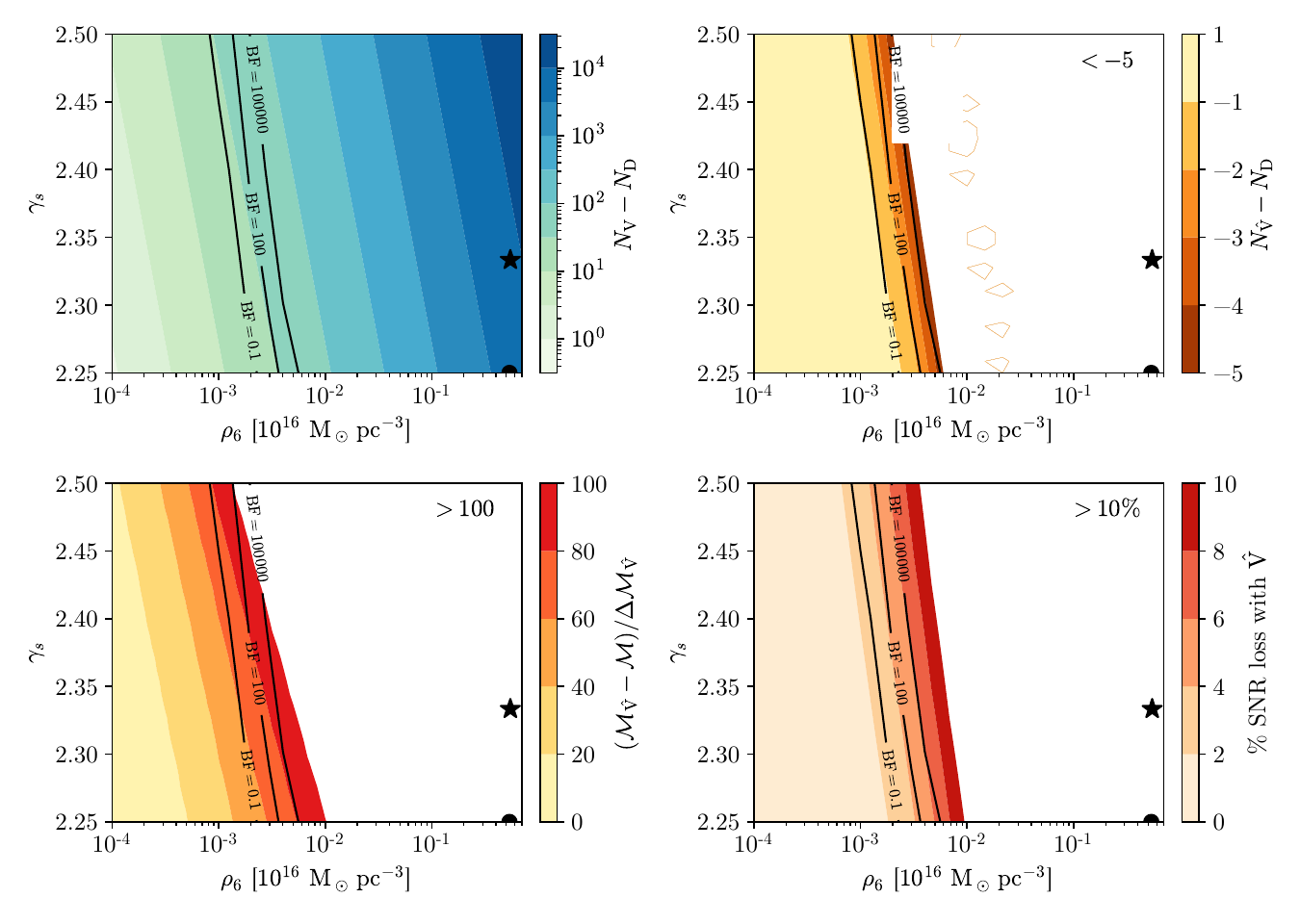}
    \caption{\textbf{Discoverability: illustration of the mismatch between dark dress and GR-in-vacuum waveforms.} In all panels the dark dress' black hole masses are fixed to $m_1 = \SI{e3}{\solarmass}$ and $m_2 = \SI{1.4}{\solarmass}$. The black contours show the Bayes factor for the dark dress vs GR-in-vacuum waveform. Upper left: number of cycles of dephasing between dark dresses and corresponding DM-free GR-in-vacuum systems. This is defined over the frequency range starting five years before the dark dress merges at its ISCO frequency. The $\hat{\mathrm{V}}$ in the other panels refers to the maximum-likelihood (ML) GR-in-vacuum systems. Upper right: dephasing between dark dresses and ML vacuum system. Lower left: ML vacuum system (detector-frame) chirp mass bias, in units of posterior width. Lower right: decrease in signal-to-noise ratio from using a GR-in-vacuum waveform to search for a dark dress. In the white regions the computations become numerically challenging. The black star ($\boldsymbol\star$) and dot ($\bullet$) indicate the astrophysical and primordial black hole benchmarks, respectively.}
    \label{fig:discoverability}
\end{figure*}

Dark dresses out to $\sim\SI{75}{\mega\parsec}$ with chirp masses above $\sim\SI{16}{\solarmass}$ would be detectable by LISA. While we assume a five-year observing window immediately preceding merger, heavier systems are detectable at this distance even earlier in their inspirals. For example, a system with component masses \SI{e5}{\solarmass} and \SI{100}{\solarmass} would have SNR higher than 15 if observed during any five year window within \SI{100}{\year} of coalescence. Additionally this detection horizon easily encompasses the Virgo Supercluster and the larger Laniakea Supercluster~\cite{Tully:2014gfa}, which contains $\sim\SI{e17}{\solarmass}$ of matter. This suggests ample opportunities for detecting signals, but converting our results into an event rate is difficult.

Estimating the detection rate requires understanding how often and at what redshifts IMBHs capture lighter companions. The formation rate of IMRI systems can be predicted for different populations of IMBHs, depending on their origin (see e.g.~Ref.~\cite{Fragione:2017blf} concerning IMBHs in MW globular clusters). Though current cosmic microwave background constraints are consistent with several million dressed \SI{e3}{\solarmass} PBHs~\cite{Serpico:2020ehh} within the detection horizon, the formation of IMRIs from dressed PBHs has not been well studied. Furthermore, the abundance of dressed astrophysical IMBHs is not well-understood. We leave these detailed population-level studies for future work.

\Cref{fig:discoverability} quantifies when a system can be discovered to be a dark dress rather than a GR-in-vacuum binary. We focus on the benchmark masses $m_1 = \SI{e3}{\solarmass}$ and $m_2 = \SI{1.4}{\solarmass}$ and vary the DM spike parameters. The nearly-vertical black contours show the Bayes factor. This demonstrates that systems with density normalizations larger than $\rho_6 = \SI{e14}{\solarmass/\parsec^3}$ could be decisively distinguished from GR-in-vacuum binaries.

By several other metrics shown in \cref{fig:discoverability}, dark dresses with density normalizations above this contour look significantly different from GR-in-vacuum ones. In the top right and bottom panels, we consider the scenario in which LISA measures a dark dress signal, but only has a template bank of GR-in-vacuum waveforms. We denote the best-fitting GR-in-vacuum system with $\hat{\mathrm{V}}$ and illustrate how much it differs from the actual dark dress system. We stop the calculations in the white regions to the right of $\rho_6 \sim \SI{e14}{\solarmass/\parsec^3}$ where it becomes difficult to optimize the vacuum system's chirp mass since the dark dress waveform is so dephased. The top right panel shows the amount of dephasing for this system over a fixed frequency range spanning from a frequency corresponding to five years before the dark dress merges to the system's ISCO frequency. For Bayes factors above 100, the dephasing is just a few cycles. In this same region, the SNR loss becomes more substantial, exceeding $10\%$ (bottom right panel). The bottom left panel illustrates the bias of this system's (detector-frame) chirp mass relative to the true value for the dark dress. This is larger than the measurement error for the chirp mass by a factor of more than 75. For reference, we also plot the dephasing between the dark dress and the same system without dark matter in the top left panel.

\begin{figure*}
    \centering
    \includegraphics[width=0.9\textwidth]{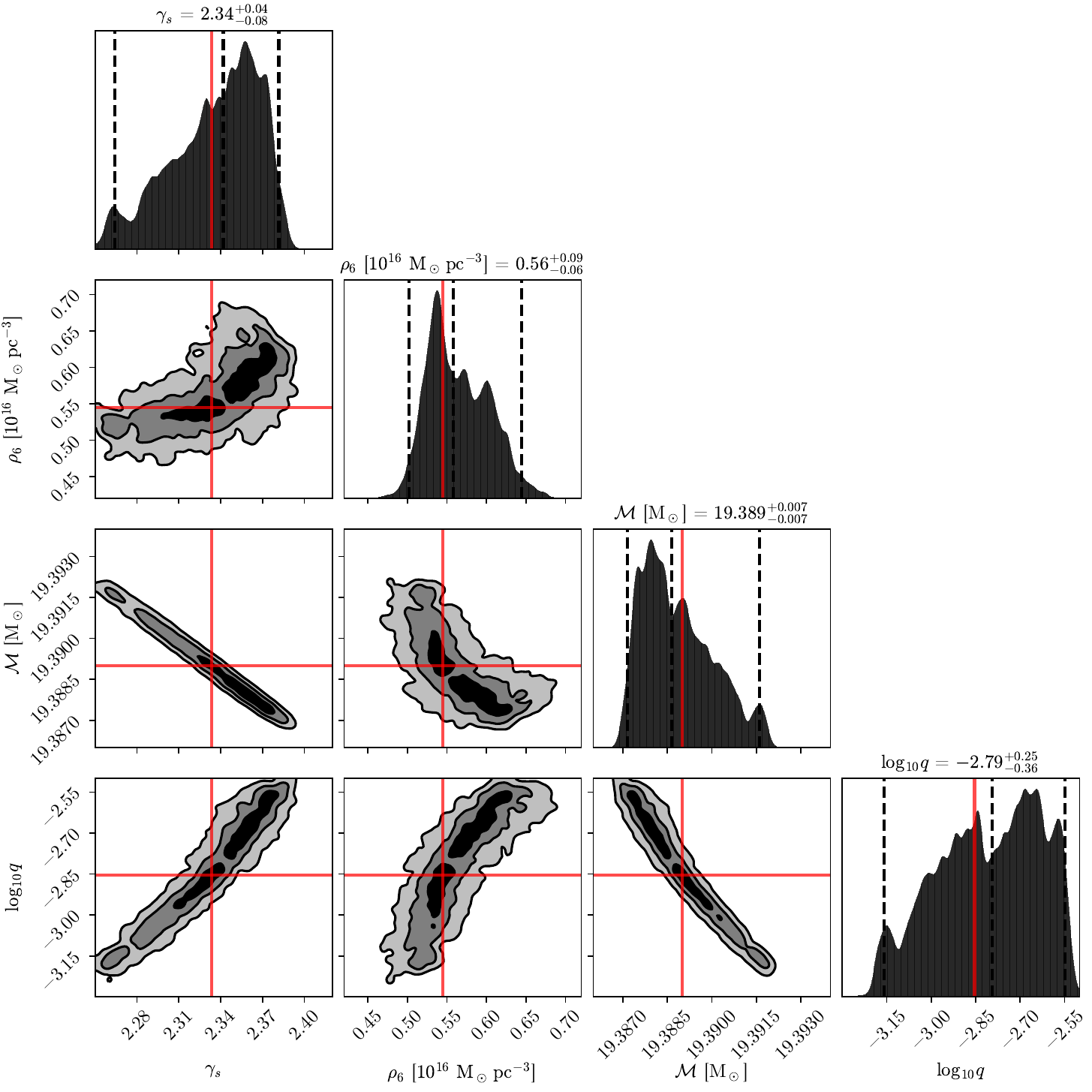}
    \caption{\textbf{Measurability: marginal posteriors for intrinsic parameters for the astrophysical dark dress benchmark (first row in \cref{tab:benchmarks}).} The red lines indicate the true parameter values, with the chirp mass defined in the detector frame. The 2D contours show the 68\%, 95\% and 99.7\% credible regions. The dashed vertical lines overlaying the 1D marginal posteriors indicate the $95\%$ credible interval and median. All posteriors have been smoothed by 1.5\% with a Gaussian kernel. Note that the parameter ranges used here are narrower than those used for the Bayes factor calculations in \cref{fig:discoverability}. \revision{The seeming multimodality is a consequence of the tight correlations in the posteriors which the nested sampler has trouble resolving, as well as numerical noise due to the difficulty of evaluating the match integral in the likelihood.}}
    \label{fig:astro-post-ns}
\end{figure*}

\begin{figure*}
    \centering
    \includegraphics[width=0.9\textwidth]{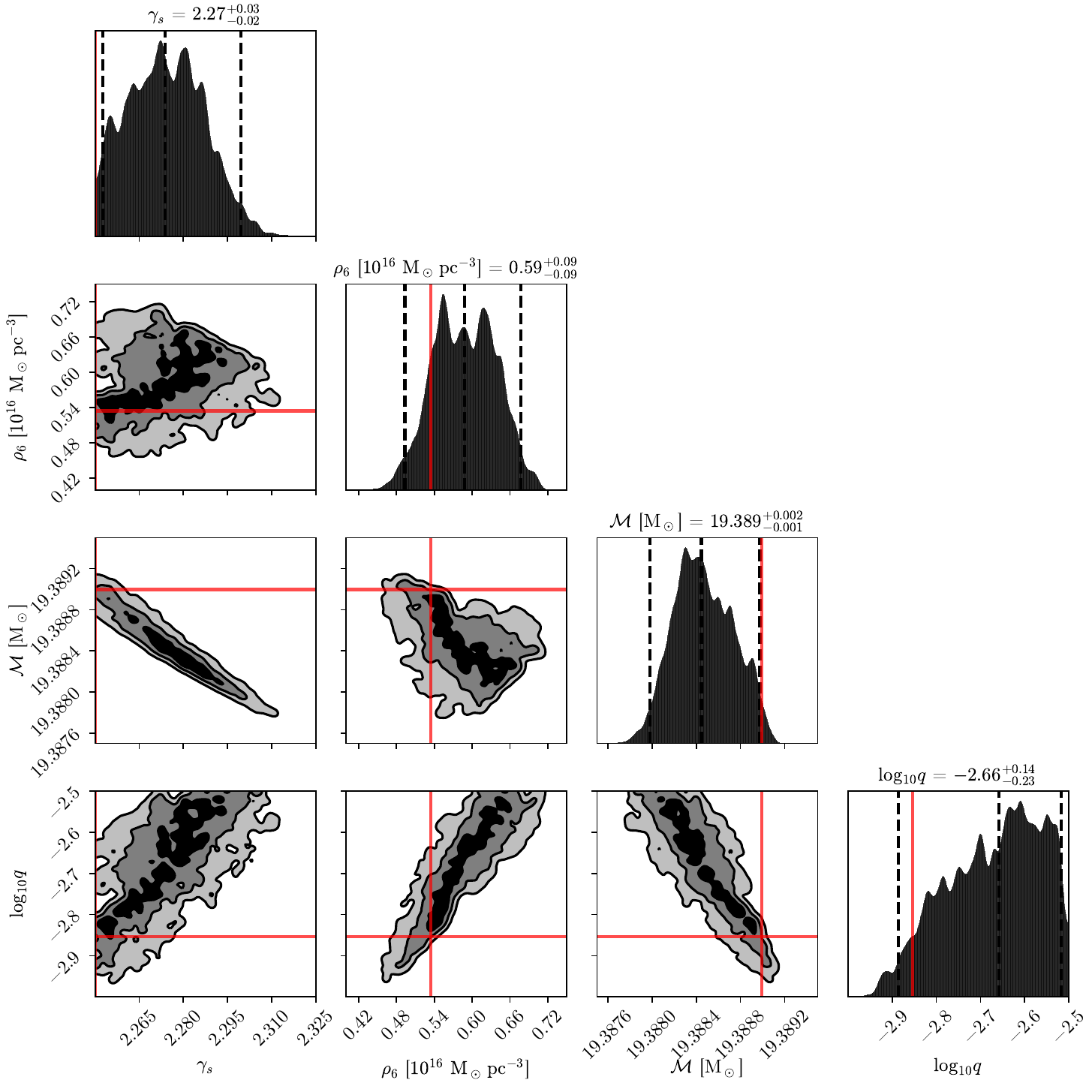}
    \caption{\textbf{Measurability: marginal posteriors for intrinsic parameters for the primordial black hole dark dress benchmark (second row in \cref{tab:benchmarks}).} See the caption of \cref{fig:astro-post-ns} for details.}
    \label{fig:pbh-post-ns}
\end{figure*}

The 1D and 2D marginal posteriors for the astrophysical and primordial black hole benchmarks are shown in \cref{fig:astro-post-ns} and \cref{fig:pbh-post-ns}, respectively. For both systems the DM halo's density normalization can be distinguished from zero at high significance, with $\rho_6 \approx 5.6_{-0.6}^{+0.9}\times10^{15}\, \si{\solarmass/\parsec^3}$ ($95\%$ credible interval) for the astrophysical benchmark and a similar level of precision in the primordial formation scenario.  This corresponds to measuring the presence of an initial $\sim\SIrange{0.08}{0.14}{\solarmass}$ of dark matter within $\SI{e-6}{\parsec}$ of the central black hole for both systems. The DM halo's slope can also be measured, albeit with large error bars due to strong degeneracies with the (detector-frame) chirp mass $\mathcal{M}$ and the mass ratio $\log_{10} q$, particularly for the astrophysical benchmark. Even so, the 1\%-3\% uncertainties on $\gamma_\rmsp$ are sufficiently small that the two benchmark values $\gamma = 2.\overline{3}$ and $\gamma = 2.25$ can be distinguished, suggesting that such measurements of the dephasing can hint at the formation mechanism of the dark dress.

The posteriors exhibit some bias, particularly for the primordial black hole benchmark where $\gamma_\rmsp$ lies on the boundary of the prior range. This is due to the relatively tight priors used for $\gamma_\rmsp$ and $\log_{10} q$. While we could permit a wider range of values for the density profile slope or consider smaller mass ratios, biases would still remain because our numerical modeling of dark dresses is unreliable for mass ratios above $\sim 10^{-2.5} \sim 0.003$.

Unlike in the GR-in-vacuum case, the mass ratio can be measured even in the Newtonian limit. This is because it enters in the frequency scales $f_b$ and $f_\mathrm{eq}$ [\cref{eq:f_b_scaling,eq:c_f}]. Though it falls beyond the scope of this paper, modeling the gravitational wave emission at first post-Newtonian order would significantly improve the measurement error for the mass ratio. The mass ratio can be measured more precisely for the primordial black hole benchmark due to the degeneracy between $\log_{10} q$ and $\gamma_\rmsp$. This is because $\gamma_\rmsp = 9/4$ lies on the prior boundary, truncating the $\gamma_\rmsp < 9/4$ part of the posterior. An additional consequence of this truncation is that we obtain smaller errors on the chirp mass than for the astrophysical benchmark, though in both cases the fractional error is $\order{\num{e-4}}$. For the corresponding DM-free GR-in-vacuum binary the errors on the chirp mass are about two orders of magnitude smaller, since there are no other intrinsic parameters with which to be degenerate.

\revision{The correlations in the posteriors can differ in sign from the analysis where the halo's evolution is neglected. For example, in the static case $\rho_6$ and $\log_{10} q$ are \emph{anticorrelated}, since increases in either quantity lead drive up the dynamical friction term in the compact object's equation of motion (c.f. \cref{eq:r_eom}). However, in our analysis increasing $\log_{10} q$ increases the dynamic dress dephasing break frequency $f_b$, which in turn \emph{decreases} the effective density profile at fixed binary separation. The effect is the \emph{correlation} seen in the $(\rho_6, \log_{10} q)$ marginal posterior. Similar reasoning explains the other relationships observed in \cref{fig:astro-post-ns,fig:pbh-post-ns}.}

\section{Discussion and Conclusions}
\label{sec:conclusions}

In this work, we studied the prospects for detecting and characterizing dark matter overdensities around intermediate mass-ratio inspirals with LISA. We introduced a new analytical approximation for gravitational waveforms from systems with a dark dress with an evolving dark matter distribution, and we validated the approximate waveforms against waveforms from full numerical simulations. We then studied the detectability (signal-to-noise ratio), and presented a Bayesian framework to assess the discoverability (discrimination against in-vacuum inspiral) and measurability (prospects for measuring dark dress parameters), assuming a detection with LISA. Our key conclusions were:
\begin{itemize}
    \item \textbf{Detectability (\cref{fig:snrs}).} The dark matter halo has little impact on the SNR of dark dresses. Systems with chirp masses larger than $\mathcal{M} \sim \SI{16}{\solarmass}$, corresponding to $(m_1, m_2) = (\SI{e3}{\solarmass}, \SI{1.4}{\solarmass})$, are detectable to distances of $D_L \sim \SI{75}{\mega\parsec}$.
    \item \textbf{Discoverability (\cref{fig:discoverability}).} Fiducial astrophysical and primordial black hole dark dresses can easily be discriminated from GR-in-vacuum systems. Not accounting for the presence of their dark matter halos would lead to overlooking their signals or extremely biased chirp mass inferences.
    \item \textbf{Measurability (\cref{fig:astro-post-ns,fig:pbh-post-ns}).} In the case of a detection, for both astrophysical and primordial black hole dark dresses, the halo's initial density normalization can be measured with $\sim 15\%$ errors, and distinguished from zero at high significance. The halo's slope can also be measured with $\lesssim 3\%$ errors, although it exhibits strong degeneracies with the chirp mass and mass ratio. Furthermore, in constrast with GR-in-vacuum inspirals, the mass ratio can be measured even in the Newtonian limit, albeit with large error bars.
\end{itemize}

We made a number of simplifying assumptions in modeling the evolution of the binary system and the DM spike. An important first caveat is that we considered a Newtonian description of the system throughout this work. However, this approximation does not affect several of our results for the following reasons: (i) we are neglecting post-Newtonian (PN) effects for systems with and without DM, so the difference in phase accumulated is not largely affected; (ii) the dephasing is predominantly accumulated at large binary separation where PN effects are small; and (iii) dynamical friction corresponds to a \textit{negative} PN-order effect for circular orbits, so it will not be confused with standard PN corrections. 
We also assumed that the DM halo is spherically symmetric and isotropic. However, the binary is \textit{not} spherically symmetric, so we eventually expect this description to break down. In particular, there should be a transfer of angular momentum from the binary to the dark matter spike. We argue in Paper I that this effect is small, and it goes in the direction of making the dephasing larger, due to the decreased relative velocity (and therefore increased dynamical friction) between the compact object and co-rotating dark matter particles. We conclude that the numerical modeling presented here is conservative, and corrections due to angular momentum injection are higher order.

We have focused on the final 5 years of the inspiral, having in mind a 5-year LISA mission. Of course, there is no guarantee that the merger event described here will occur during the LISA observation period. If the system were to be observed at a much earlier stage, the signal-to-noise ratio and the amount of dephasing could differ significantly. 

In addition to improving models of dark dress evolution, translating our detectability results into predictions for the event rate at LISA requires further astrophysical modeling. Key inputs are the number of intermediate-mass black holes expected to be enclosed in dark matter halos, the fraction of these systems that survive to low redshifts and how often they form binaries with lower-mass companions. Applying dark dress modeling to data will further require new analysis techniques for LISA data. Interpreting future results will also depend on \revision{the relative importance of other environmental effects. For instance,} understanding how the dephasing induced by dark dresses differs from the dephasing of other systems, such as inspirals with accretion disks~\cite{Toubiana:2020drf}. \revision{The effect of mass accretion onto the smaller orbiting object should also be assessed carefully. Previous studies based on the assumption of a static DM spike \cite{Macedo:2013qea,Yue:2017iwc,Cardoso:2019rou} agreed that this process is inconspicuous compared to friction. A more accurate study of this effect in presence of halo feedback is worth considering in the future.}

We have argued that PN effects and more accurate modeling of the DM halo should not substantially affect the detectability of dark dress systems. However, these effects will be essential to include in the waveform modeling for the final LISA analysis. This is because LISA may be able to discover the presence of a dark dress with only a few cycles of dephasing (as illustrated in the upper-left panel of \cref{fig:discoverability}). Given that the $\SI{5}{\year}$ inspiral typically consists of millions of GW cycles, accurate characterization of these systems may require modeling with precision at the level of 1 part in $10^6$. Future work will therefore require PN effects to be incorporated, as well as generalizing the analysis to include eccentric orbits and the evolution of angular momentum in the DM halo. Similarly, the approximate dephasing formalism which we developed in \cref{sec:ana_model} will not be accurate enough for real data analysis. However, with this formalism we have been able to demonstrate that dark dress systems should be discoverable and measurable, motivating further work in this direction.

In conclusion, this work provides an important step towards realistic modeling of the inspiral of stellar-mass compact objects around intermediate-mass black holes surrounded by dark matter halos. It enables rapid, approximate calculation of these systems' gravitational waveforms, and shows that gravitational wave detectors could characterize their dark matter overdensities. Detecting dark dresses would have an impact beyond astrophysics and cosmology since their density profiles depend on the dark matter's fundamental properties. Measuring their dephasing would therefore provide a powerful probe of the particle nature of dark matter.

\acknowledgments

We thank Thomas Edwards and Sara Algeri for helpful discussions. We also thank Niklas Becker for catching an error in \cref{eq:rho6-def} in the first version of this work.

A.C.\ is partially funded by the Netherlands eScience Center (grant number ETEC.2019.018) and the Schmidt Futures foundation. 
D.G.\ has received financial support through the Postdoctoral Junior Leader Fellowship Programme from la Caixa Banking Foundation (grant n.~LCF/BQ/LI18/11630014).
D.G.\ was also supported by the Spanish Agencia Estatal de Investigaci\'{o}n through the grants PGC2018-095161-B-I00, IFT Centro de Excelencia Severo Ochoa SEV-2016-0597, and Red Consolider MultiDark FPA2017-90566-REDC.
D.G. also acknowledges funding from the ``Department of Excellence'' grant awarded by the Italian Ministry of Education, University and Research (MIUR).
D.G. also acknowledges support from the INFN grant ``LINDARK'', and the project ``Theoretical Astroparticle Physics (TAsP)'' funded by the INFN.
D.G. also acknowledges the support from Generalitat Valenciana through the plan GenT program (CIDEGENT/2021/017).
B.J.K.\ thanks the Spanish Agencia Estatal de Investigaci\'on (AEI, Ministerio de Ciencia, Innovación y Universidades) for the support to the Unidad de Excelencia Mar\'ia de Maeztu Instituto de F\'isica de Cantabria, ref. MDM-2017-0765.
D.A.N.\ acknowledges support from the NSF Grant No.\ PHY-2011784.

This work used the Lisa Compute Cluster at SURFsara, which runs on 100\% wind energy. We used the following software: python, \texttt{jax}~\cite{jax2018github}, \texttt{numpy}~\cite{numpy}, \texttt{scipy}~\cite{scipy}, \texttt{matplotlib}~\cite{Hunter:2007}, \texttt{jupyter}~\cite{jupyter} and \texttt{tqdm}~\cite{tqdm}.

\onecolumngrid

\begin{appendix}

\section{Deriving the break frequency scaling relation}
\label{app:f_b}

The GW phase of a dynamic dark dress system can be modeled as a broken power law in the GW frequency $f$. The scaling of the break frequency $f_b$ can be obtained by setting equal the timescale for inspiralling due to gravitational wave emission $t_\mathrm{GW}(r_2)$ and the timescale for depletion of the DM halo $t_\mathrm{dep}$ as a function of orbital radius $r_2$.

Neglecting the contribution of dynamical friction to the orbital evolution, we can write:
\begin{equation}
    \dot{r}_2 = - \frac{64\, G_N^3\, M \, m_1\, m_2}{5\, c^5\, (r_2)^3} \, .
\end{equation}
The GW timescale can then be written straightforwardly as:
\begin{equation}
    t_\mathrm{GW} \sim r_2/\dot{r}_2 \sim \frac{5 c^{5}\left(r_{2}\right)^{4}}{64 G_N^{3} (m_1 + m_2) m_{1} m_{2}} \propto (r_2)^4 m_1{}^{-2} m_2{}^{-1}\,,
\end{equation}
where we assume $m_1 \gg m_2$.

In order to derive the depletion timescale for DM particles at a given radius $r_2$, we first consider the depletion of particles with a given energy $\mathcal{E}$:
\begin{equation}
    t_\mathrm{dep}(\mathcal{E}) \sim f(\mathcal{E},t) \left|\frac{\partial f(\calE, t)}{\partial t}\right|^{-1}\,.
\end{equation}
The full time-evolution of $f(\mathcal{E}, t)$ is given in \cref{eq:dfdt} but here we will take a simpler approach and neglect the second term on the right-hand side of \cref{eq:dfdt}, which corresponds to replenishment of DM particles scattered from $\mathcal{E} - \Delta \mathcal{E} \rightarrow \mathcal{E}$. With this simplification, we can write:
\begin{equation}
    t_\mathrm{dep}(\mathcal{E}) \sim \frac{T_\mathrm{orb}}{p_\mathcal{E}}\,.
\end{equation}
Here, $p_\mathcal{E}$ is the probability that a particle with energy $\mathcal{E}$ will scatter during one orbit of the compact object and so $T_\mathrm{orb}/p_\mathcal{E}$ is the typical timescale between scatters. However, a DM particle is not completely unbound with a single ``kick'' from the compact object, but instead increases its energy by a typical amount $\langle \Delta \mathcal{E} \rangle$. Only particles moving slower than the orbiting object are considered relevant for dynamical friction, so a number of kicks $N_\mathrm{req}$ are required to increase the speed of the particle from $v$ to $v_\mathrm{orb}(r_2)$. This corresponds to a change in the relative specific energy from $\mathcal{E}$ to $\frac{1}{2}\Psi(r_2) \sim v_\mathrm{orb}(r_2)^2$. The typical number of kicks required is then:
\begin{equation}
    N_\mathrm{req} \sim \frac{\mathcal{E} - \frac{1}{2}\Psi(r_2)}{\langle \Delta \mathcal{E}\rangle}\,,
\end{equation}
and the relevant depletion timescale is:
\begin{equation}
    t_\mathrm{dep}(\mathcal{E}) \sim N_\mathrm{req} \frac{T_\mathrm{orb}}{p_\mathcal{E}}\,.
\end{equation}

The per-orbit scattering probability is given by:
\begin{equation}
    p_{\mathcal{E}} = \int P_{\mathcal{E}}(\Delta \mathcal{E}) \mathrm{d} \Delta \mathcal{E}\,,
\end{equation}
where the differential probability $P_{\mathcal{E}}(\Delta \mathcal{E})$ is given in Eq.~(4.15) of Paper I~\cite{Kavanagh:2020cfn} as:
\begin{equation}
    P_{\mathcal{E}}(\Delta \mathcal{E}) =\frac{4 \pi^{2} r_{2}}{g(\mathcal{E})} \frac{b_{90}^{2}}{v_{0}^{2}}\left[1+\frac{b_{\star}^{2}}{b_{90}^{2}}\right]^{2} \int \sqrt{2\left(\Psi\left(r\left[b_{\star}, \alpha\right]\right)-\mathcal{E}\right)} \sin \left(\theta\left[b_{\star}, \alpha\right]\right) \mathrm{d} \alpha\,.
\end{equation}
Here, $v_0 \equiv v_\mathrm{orb}(r_2) \approx \sqrt{G_N m_1/r_2}$ for brevity, while $b_{90}$ is the impact parameter corresponding to a $90^\circ$ deflection,
\begin{equation}
    b_\mathrm{90} = \frac{G_N m_2}{v_0^2}\,,
\end{equation}
and $b_\star$ is the impact parameter corresponding to kick of size $\Delta \mathcal{E}$:
\begin{equation}
    b_* = b_{90} \sqrt{\frac{2 v_0^2}{|\Delta\mathcal{E}|} - 1}\,.
\end{equation}
We will take the integral over the angular variable $\alpha$ to be of order 1, meaning that $P_{\mathcal{E}}(\Delta \mathcal{E})$ can be re-expressed as:
\begin{align}
    P_{\mathcal{E}}(\Delta \mathcal{E}) &\approx \frac{16 \pi^2 r_2}{g(\mathcal{E})} \frac{G_N^2 m_2^2}{v_0^2 \Delta \mathcal{E}^2} \sqrt{2\left( \Psi(r_2) - \mathcal{E} \right)}\\
    &= \frac{16}{\sqrt{2}} \left(\frac{m_2}{m_1}\right)^2 \frac{1}{v_0^4 \Delta \mathcal{E}^2} \mathcal{E}^{5/2}  \sqrt{2\left( \Psi(r_2) - \mathcal{E} \right)}\,.
\end{align}
We can perform the integral over $\Delta\mathcal{E}$ in the range
\begin{align}
    \Delta \mathcal{E}_{\min } &= 2 v_{0}^{2} \left[1+\frac{b_{\max }^{2}}{b_{90}^{2}}\right]^{-1} = 2 v_0^2 \left[ 1 + \frac{m_1}{m_2} \right]^{-1}\\
    \Delta \mathcal{E}_{\max } &= 2 v_{0}^{2}\left[1+\frac{b_{\min }^{2}}{b_{90}^{2}}\right]^{-1} \approx 2 v_0^2\,,
\end{align}
where we have used $b_\mathrm{max} = \sqrt{m_2/m_1} r$. With this, we obtain:
\begin{align}
    p_\mathcal{E} = 8\frac{m_2}{m_1} \frac{r_2^2}{G_N^3 m_1^3} \mathcal{E}^{5/2}\sqrt{\Psi(r_2) - \mathcal{E}}\,.
\end{align}
The mean kick size is
\begin{equation}
    \langle \Delta\mathcal{E} \rangle = \frac{1}{p_{\mathcal{E}}} \int \Delta\mathcal{E} \, P_{\mathcal{E}}(\Delta\mathcal{E}) \, \mathrm{d} \Delta\mathcal{E} = \frac{2 m_2 v_0^2}{m_1} \log\left[ 1 + \frac{m_1}{m_2} \right]\,.
\end{equation}
Putting everything together gives
\begin{align}
    t_\mathrm{dep}(\mathcal{E}) \sim \frac{ m_1^2 \, T_\mathrm{orb}}{32 \, m_2^2 \, \log(1 + m_1 / m_2)} \frac{2x - 1}{x^{5/2} \sqrt{1 - x}}\,, \label{eq:t_dep_e}
\end{align}
where $x = \mathcal{E} / \Psi(r_2)$ and we remind the reader that $\Psi(r_2) = G_N m_1/r_2$.

To convert from $t_\mathrm{dep}(\mathcal{E})$ to $t_\mathrm{dep}(r_2)$, we must compute an average with respect to the phase-space distribution of particles moving more slowly than the local circular speed. Using the fact that close to the central BH $f(\mathcal{E}) = \mathcal{N} \mathcal{E}^{\gamma_\rmsp - 3/2}$, this amounts to
\begin{align}
    t_\mathrm{dep}(r_2) &\sim \frac{4\pi \int_{\frac{1}{2} \Psi(r_2)}^{\Psi(r_2)} \mathrm{d}\mathcal{E}\, t_\mathrm{dep}(\mathcal{E})\, f(\mathcal{E}) \sqrt{2(\Psi(r_2) - \mathcal{E})}}{4\pi \int_{\frac{1}{2} \Psi(r_2)}^{\Psi(r_2)} \mathrm{d}\mathcal{E}\, f(\mathcal{E}) \sqrt{2(\Psi(r_2) - \mathcal{E})}}\\
    &= \frac{\int_{\frac{1}{2}}^{1} \mathrm{d}x\, t_\mathrm{dep}(\mathcal{E})\, x^{\gamma_\rmsp - 3/2} \sqrt{1 - x}}{\int_{\frac{1}{2}}^{1} \mathrm{d}x\, x^{\gamma_\rmsp - 3/2} \sqrt{1 - x}}\\
    &\equiv \frac{1}{h(\gamma_\rmsp)} \, \int_{\frac{1}{2}}^{1} \mathrm{d}x\, t_\mathrm{dep}(\mathcal{E})\, x^{\gamma_\rmsp - 3/2} \sqrt{1 - x},
\end{align}
where
\begin{equation}
    h(\gamma_\rmsp) \equiv \operatorname{B}_{1}\left(\gamma_\rmsp-\frac{1}{2}, \frac{3}{2}\right)-\operatorname{B}_{\frac{1}{2}}\left(\gamma_\rmsp-\frac{1}{2}, \frac{3}{2}\right)\,,
\end{equation}
and $B_x(a, b)$ is the incomplete beta function,
\begin{equation}
    B_{x}(a, b)=\int_{0}^{x} t^{a-1}(1-t)^{b-1} \,\mathrm{d} t\,.
\end{equation}
Substituting in for $t_\mathrm{dep}(\mathcal{E})$ with \cref{eq:t_dep_e} yields
\begin{align}
    t_\mathrm{dep}(r_2)   &= \frac{m_1^2 T_\mathrm{orb}}{32 m_2^2 \log(1 + m_1 / m_2)} \, g(\gamma_\rmsp)\\
    &= \frac{\pi \, m_1^{3/2} \, r_2^{3/2}}{16 \sqrt{G_N} \, m_2^2 \, \log(1 + m_1 / m_2)} \, g(\gamma_\rmsp)\,,
\end{align}
where we have used $T_\mathrm{orb} = 2\pi\sqrt{r_2^3 / G_N m_1}$ and defined
\begin{equation}
    g(\gamma_\rmsp) \equiv \frac{4 - \gamma_\rmsp - 2^{3-\gamma_\rmsp}}{(3-\gamma_\rmsp)(2-\gamma_\rmsp) \, h(\gamma_\rmsp)}.
\end{equation}

Equating $t_\mathrm{GW}$ and $t_\mathrm{dep}$ and solving for the break radius, we find:
\begin{equation}
    r_b = \left(\frac{4\pi }{5} \frac{m_1 g(\gamma_\rmsp)}{m_2 \log(1 + m_1 / m_2)}\right)^{2/5} \,\left(\frac{G_N m_1}{c^2}\right)\,.
\end{equation}
Using $f \approx \sqrt{G_N m_1/r_2^3}/\pi$, the corresponding break frequency is:
\begin{equation}
    f_b = \frac{1}{\pi}\sqrt{G_N m_1}\left(\frac{G_N m_1}{c^2}\right)^{-3/2} \left(\frac{4\pi }{5} \frac{m_1 g(\gamma_\rmsp)}{m_2 \log(1 + m_1 / m_2)}\right)^{-3/5} \propto m_1^{-8/5} m_2^{3/5} \left[\log(1 + m_1 / m_2)\, g(\gamma_\rmsp)\right]^{3/5}\,.
\end{equation}
For a system with $m_1 = 1000\,M_\odot$, $m_2 = 1\,M_\odot$ and $\gamma_\rmsp = 7/3$, this estimate gives $f_b \approx 1.22 \,\mathrm{Hz}$ independent of the density normalization $\rho_\rmsp$.

\section{From phase to strain}
\label{app:phase-to-strain}

Here we review how to compute the gravitational wave strain from the phase at leading Newtonian order in the binary's dynamics (see e.g.~\cite{Maggiore:2007ulw}). We will work with the two polarization modes of the gravitational waves, the plus and cross polarizations, which we will denote by $h_+$ and $h_\times$.
In the Newtonian limit the polarizations are determined by the quadrupole waves, which are related to the two polarizations by
\begin{equation}
    h_+ - i h_\times = h_{2,2}(t) {}_{-2}Y_{22}(\iota,\phi) + h_{2,-2}(t) {}_{-2}Y_{2-2}(\iota,\phi) \, ,
\end{equation}
where ${}_{-2}Y_{22}(\iota,\phi_c)$ is a spin-weighted spherical harmonic of spin weight $-2$. This leads to the quadrupole expression for the two polarizations
\begin{equation}
\begin{split}
    h_+(t) &= \frac{4 G_N \mu}{c^4 D_L} \frac{1 + \cos^2\iota}{2} (\omega r_2 )^2 
    \cos[2\Phi_\mathrm{orb}(t) + 2\phi] \, ,\\
    h_\times(t) &= \frac{4 G_N \mu}{c^4 D_L} \cos\iota  (\omega r_2 )^2 \sin[2\Phi_\mathrm{orb}(t) + 2\phi] \, .
\end{split}
\end{equation}
where $\Phi_\mathrm{orb}(t)$ is the orbital phase and $\omega = \dot \Phi_\mathrm{orb}(t)$ is the orbital frequency.

We can then take the Fourier transform of the two polarizations,
\begin{equation}
    \tilde h_{+,\times} (f) = \int_{-\infty}^\infty dt e^{i 2\pi f} h_{+,\times}(t) \, .
\end{equation}
We take stationary-phase approximation to the Fourier modes, in which the Fourier transform is evaluated using the method of steepest descent.
It is convenient to write the result in terms of an amplitude and phase as
\begin{equation}
    \tilde h_{+,\times}(f) = A_{+,\times}(f) \, e^{i \Psi(f)} \, .
\end{equation}
The phase depends on the (luminosity) distance $D_L$ to the binary as well as the phase at and time of coalescence $\phi_c$ and $t_c$:
\begin{align}
    \Psi(f) &= 2 \pi f \qty(t_c + \frac{D_L}{c} - t(f)) + \Phi(f) - \phi_c - \frac{\pi}{4} \, . \label{eq:strain_phase}
\end{align}
The two polarization amplitudes are proportional to an intrinsic piece $h_0$ and functions of the inclination angle $\iota$:\footnote{
    This is the angle between the line of sight and rotational axis of the binary.
}
\begin{align}
    A_+(f) &= \frac{1}{D_L} \frac{1 + \cos^2\iota}{2} h_0(f) \label{eq:amp_plus}\\
    A_\times(f) &= \frac{1}{D_L} \cos\iota\, h_0(f) \, , \label{eq:amp_times}\\
    h_0(f) &= \frac{1}{2} \frac{4 \pi^{2/3} G_N^{5/3} \mathcal{M}^{5/3} f^{2/3}}{c^4} \sqrt{\frac{2\pi}{\ddot{\Phi}(f)}} \, \label{eq:h0}.
\end{align}
We provide analytic expressions for $t(f)$ and $\ddot{\Phi}(f)$ for our waveform parametrization in \cref{app:waveform_exprs}. The strain measured by a detector is a linear combination of the polarizations,
\begin{equation}
    \tilde{h}(f) = F_+ \, \tilde{h}_+(f) + F_\times \, \tilde{h}_\times(f) \, ,
\end{equation}
where the detector pattern functions $F_{+,\times}$ depend in general on frequency and the location of the binary. In this work we assume the detector measures the strain averaged over inclination angle (see e.g. Ref~\cite{Robson:2018ifk}),
\begin{equation}
    \tilde{h}(f) = \sqrt{\frac{4}{5}} \frac{h_0(f)}{D_L} \, .
\end{equation}

\section{Useful expressions for computing waveforms}
\label{app:waveform_exprs}

For convenience we provide the expressions required to compute the strain phase (\cref{eq:strain_phase}) and amplitude (\cref{eq:amp_plus,eq:amp_times,eq:h0}) using our analytic approximation waveform model from \cref{eq:phase_hyp}. Using the relation
\begin{equation}
    \ddot{\Phi} = 4\pi^2 f \qty(\dv{\Phi}{f})^{-1} \, ,
\end{equation}
and recalling that $y \equiv f / f_t$, the phase acceleration is
\begin{equation}
    \ddot{\Phi}(f) = \frac{12 \pi^2 f^{11/3}}{a_V} \qty{ 5 - \eta y^{-\lambda} \qty[ 3 \lambda + 5\qty(1+y^{\frac{5}{3 \vartheta}})^{-1} - 3 \lambda\, \hyp\qty(1, \vartheta, 1 + \vartheta, -y^{-\frac{5}{3 \vartheta}}) ] }^{-1} \, ,
\end{equation}
where we defined $a_V \equiv \frac{1}{16} \qty(\frac{c^3}{\pi G_N \mathcal{M}})^{5/3}$. The time elapse since an arbitrary initial frequency is given by
\begin{equation}
\begin{split}
    t(f) &= \frac{1}{2\pi} \int \frac{\dd{f'}}{f'} \dv{\Phi}{f}\\
    &= \frac{a_V y^{-\lambda}}{16\pi (1 + \lambda) (8 + 3 \lambda) f^{8/3}} \Biggl\{
        5 (1 + \lambda) (8 + 3 \lambda) y^\lambda
        + 8 \lambda (8 + 3 \lambda) \eta \hyp\qty( 1, \vartheta, 1 + \vartheta, -y^{-\frac{5}{3 \vartheta}} ) \\
    &\hspace{5cm}
        - 40 (1 + \lambda) \eta \hyp\qty( 1, -\frac{\vartheta (8 + 3 \lambda)}{5}, 1 - \frac{\vartheta (8 + 3\lambda)}{5}, -y^{\frac{5}{3 \vartheta}} ) \\
    &\hspace{5cm} - 8 \lambda \eta \qty[ 3 + 3 \lambda + 5 \hyp\qty( 1, \frac{\vartheta (8 + 3 \lambda)}{5}, 1 + \frac{\vartheta (8 + 3 \lambda)}{5}, -y^{-\frac{5}{3 \vartheta}} ) ]
    \Biggr\} \, .
\end{split}
\end{equation}
For a static dress these expression simplify substantially to
\begin{align}
    \ddot{\Phi}(f) &= \frac{12 \pi^2 \qty( f^{11/3} + c_f \, f^{2 \gamma_\rmsp / 3} )}{5 a_V}\\
    t(f) &= \frac{5 a_V}{16 \pi f^{8/3}}\, {}_2F_1\qty[ 1, \frac{8}{11 - 2 \gamma_\rmsp}, 1 + \frac{8}{11-2\gamma_\rmsp}, -c_f\, f^{\frac{2\gamma_\rmsp - 11}{3}} ] \, .
\end{align}
Finally, in the vacuum case
\begin{equation}
    \ddot{\Phi}(f) = \frac{12 \pi^2 f^{11/3}}{5 a_V} \, , \quad t(f) = \frac{5 a_V}{16 \pi f^{8/3}} \, .
\end{equation}

\end{appendix}

\bibliography{references}

\end{document}